\documentclass[sigconf,authorversion,nonacm]{acmart}

\def\BibTeX{{\rm B\kern-.05em{\sc i\kern-.025em b}\kern-.08em
    T\kern-.1667em\lower.7ex\hbox{E}\kern-.125emX}}

\usepackage{booktabs} % For formal tables

\usepackage[ruled,linesnumbered]{algorithm2e}
\usepackage{algorithmic}
\usepackage{graphicx}
\usepackage{textcomp}
\usepackage{xcolor}
\usepackage{url}
\usepackage{tabularx,ragged2e,booktabs,caption}

\usepackage{times,amsmath,epsfig}
\usepackage{enumerate}
\usepackage{subfigure}
\usepackage{multirow}
\usepackage{tabularx, subfigure}
\usepackage{hyperref}
\usepackage{float} %fix the position of figures and table by adding [H]
\usepackage{array}
\usepackage{makecell}

\usepackage{color}

\newcommand{\Change}[1]{{\bf\color{red} #1}}

\newcommand{\eat}[1]{}

\begin{document}
\title{Efficient Dynamic Clustering: Capturing Patterns from Historical Cluster Evolution}
% \titlenote{Produces the permission block, and copyright information}
% \subtitle{Extended Abstract}
% \subtitlenote{The full version of the author's guide is available as
%   \texttt{acmart.pdf} document}

\author{Binbin Gu}
\affiliation{%
  \institution{Unversity of California, Irvine}
  \streetaddress{}
  %\city{Irvine}
  %\state{CA}
  \postcode{}
}
\email{binbing@uci.edu}

\author{Saeed Kargar}
\orcid{}
\affiliation{%
  \institution{Unversity of California, Santa Cruz}
  \streetaddress{}
  %\city{Santa Cruz}
  %\state{CA}
}
\email{skargar@ucsc.edu}

\author{Faisal Nawab}
\orcid{}
\affiliation{%
  \institution{Unversity of California, Irvine}
 %\city{Irvine}
  %\state{CA}
}
\email{nawabf@uci.edu}

% \renewcommand{\shortauthors}{B. Trovato et al.}
\renewcommand{\shortauthors}{}

\begin{abstract}
Clustering aims to group unlabeled objects based on similarity inherent among them into clusters. It is important for many tasks such as anomaly detection, database sharding, record linkage, and others. Some clustering methods are taken as batch algorithms that incur a high overhead as they cluster all the objects in the database from scratch or assume an incremental workload.
In practice, database objects are updated, added, and removed from databases continuously which makes previous results stale. Running batch algorithms is infeasible in such scenarios as it would incur a significant overhead if performed continuously. This is particularly the case for high-velocity scenarios such as ones in Internet of Things applications.

In this paper, we tackle the problem of clustering in high-velocity \emph{dynamic scenarios}, where the objects are continuously updated, inserted, and deleted. Specifically, we propose a generally dynamic approach to clustering that utilizes previous clustering results. Our system, DynamicC, uses a machine learning model that is augmented with an existing batch algorithm. The DynamicC model trains by observing the clustering decisions made by the batch algorithm. After training, the DynamicC model is used in cooperation with the batch algorithm to achieve both accurate and fast clustering decisions. 
The experimental results on four real-world and one synthetic datasets show that our approach has a better performance compared to the state-of-the-art method while achieving similarly accurate clustering results to the baseline batch algorithm.
\end{abstract}

\maketitle

\section{Introduction}

Clustering is an unsupervised learning method which groups the objects  based on some similarity inherent among them into clusters. 
The clustering problems, such as k-means \cite{macqueen1967some}, DBSCAN \cite{ester1996density} and correlation clustering \cite{bansal2004correlation}, have been addressed in many contexts such as anomaly detection \cite{bailis2017macrobase}, database sharding \cite{curino2010schism}, record linkage \cite{gruenheid2014incremental}, recommendation systems \cite{gong2010collaborative} and others.
%
% Some clustering algorithms that take the database as input and cluster its data according to a similarity function that optimizes metrics such as Correlation Clustering \cite{bansal2004correlation} and DB-index \cite{gruenheid2014incremental}. %
%
Although a lot of clustering methods have been proposed and studied (surveyed in \cite{berkhin2006survey, rai2010survey}), some of them cluster data from scratch---which we will call the \emph{batching approach} hereafter---making them incur a high overhead.
%for getting good clustering results.

In practice, databases are dynamic. The data in the databases is continuously updated, added and removed from databases making previous clustering results stale. This makes the batching approach unsuitable for these workloads as they incur significant overhead. This is especially the case for high-velocity workloads in scenarios such as Internet of Things (IoT)
%, \Change{e.g. TIPPERS~\footnote{https://tippers.ics.uci.edu/}}, 
and large-scale Internet and cloud applications.

Although some dynamic approaches have been proposed for specific clustering methods, these approaches are not general and may incur high time complexity. For instance, Gan et. al. \cite{gan2017dynamic} propose a dynamic algorithm which is customized to the specific density-based method $\rho$-approximate DBSCAN. 
%
% Consequently, incremental strategies for clustering have been introduced in~\cite{gruenheid2014incremental, whang2014incremental} to overcome the limitations of the batching approach.
%
Gruenheid et. al. \cite{gruenheid2014incremental} and Whang et. al. \cite{whang2014incremental} focus on finding properties of the specific algorithms that could be leveraged to make lightweight changes to clustering without having to run a batch algorithm from scratch. The main limitation of these approaches is that each variant relies on specific properties and assumptions of a certain clustering algorithm, such as locality and monotonicity. This makes a proposed incremental method not applicable to other instances of the corresponding clustering algorithm and certainly not applicable to other clustering algorithms.

In this work, we make the observation that a batch clustering algorithm makes clustering changes to a specific workload in predictable patterns. Specifically, the pattern of how an existing clustering changes in reaction to an add, delete, or modify operation, can be learned to predict future changes without involving the original batch clustering algorithm. 
To this end, we propose DynamicC, a machine learning-based system that is augmented with an existing batch clustering algorithm. DynamicC undergoes a training phase, where it observes the cluster evolution patterns while using a batch clustering algorithm. Then, DynamicC is used---instead of the original batch clustering algorithm---to make clustering decisions in reaction to data operations.
The benefits of using DynamicC---compared to the original batching algorithm---is that it can be smaller and faster because it learns the clustering patterns for a specific workload. Additionally, DynamicC dynamically learns the cluster evolution patterns without making stringent assumptions about the used batching algorithm. Therefore, DynamicC can be augmented with a wide-range of clustering algorithms with minimal changes unlike existing incremental methods.

The main challenges we tackle in this work is the design of the training and prediction phases of DynamicC. Specifically, it is not trivial to represent cluster evolutions because the space of information that are needed to make decisions and the space of decisions are large. The space of information is the similarity relations between all nodes, and the decision space is all the possible clustering permutations of these nodes. In the paper, we present the details of how we represent the clustering information and decisions efficiently and compactly while enabling us to make accurate predictions. At a high level, we represent clustering information as the changes applied to the database in addition to global similarity metrics. Also, we represent cluster changes in terms of binary decisions on whether a cluster would merge or split in reaction to the new changes on the database. 

The compact information and decision representation in DynamicC enables tractable training and prediction of the machine learning model. However, it also leads to the need to do more processing to interpret the general decisions (merge or split a cluster) to specific actions (which nodes should move from one cluster to another.) To this end, we propose heuristics to translate general decision from the machine learning model to specific actions to be applied to clusters during the prediction phase. These heuristics use the clustering objective function to make the decisions of which nodes should leave or join specific clusters. A nice feature of this heuristic is that using the objective function enables us to know whether the predicted change is accurate or erroneous (if the change leads to a lower objective function score, then we know that the prediction is not accurate and we avoid it.)

In the long run, DynamicC can adjust its machine learning model by retraining continuously on new objects and observing the erroneous predictions during operation. This can be performed while running the original batching algorithm occasionally to establish a baseline for accuracy.

% \Binbin{Maybe we should first put an example here and then rephrase this paragraph.}
% At a high level, the ML model is taken as a guide which tells us which clusters tend to change. In this way, the search space is dramatically reduced especially when the algorithm is complicated (i.e. with high time complexity). Instead of focusing on improving the accuracy of the ML model, we propose a way that can accurately find those candidate clusters i.e. the clusters that will change when new updates come. We show that finding all the positive clusters can be done on the training data which suggests that we can make the recall of the model reach $100\%$ on the training data. However, training  such  a  model  is  not  straight-forward with machine learning techniques and degenerate the ML model. Therefore, we propose an effective way to predict new items instead of changing the original ML model.

To summarize, DynamicC addresses the the dynamic clustering problem using a ML model that can be augmented and trained with a wide-range of batching algorithms. This overcomes the challenges of prior work that are either batch algorithms that are too expensive for high-velocity dynamic workloads, or incremental algorithms that are restricted to specific clustering algorithms and cannot be extended to other algorithms.
Our experiments on real-world and synthetic datasets show that our method can be 85\% faster than the state-of-the-art method while retaining accurate clustering results that are within 2\% (in terms of F1) of results from the baseline batching algorithm.

The rest of the paper is organized as follows: We first present an overview and related work in Section 2. Then, we state our problem formally and present relevant background in Section 3. We introduce DynamicC's method to monitor and capture historical clustering evolution in Section 4. We propose DynamicC's ML model that learns the patterns from the captured historical cluster evolution in Section 5. In Section 6, we introduce the DynamicC algorithms that use the ML model. The experimental evaluation is covered in Section 7. We conclude in Section 8.

\section{Overview and Related Work}

\subsection{Overview}

\begin{figure}[!ht]
\setlength{\abovecaptionskip}{0.cm}
\setlength{\belowcaptionskip}{-0.cm}
\centering
% \subfigure[]{
%     \begin{minipage}[h]{0.45\textwidth}
%     \centering
%     \renewcommand\arraystretch{1.2}
%     \begin{tabular}{c|c|c|c} 
%          & {\bf Name} & {\bf Address} & {\bf Phone} \\ \hline
%         $r_1$ & Smith John & 100 MISSION ST & 8004311928\\ \hline
%         $r_2$ & S. John & 124 Mission Street & 8312907398\\ \hline
%         $r_3$ & Shari John & 124 Mission Street & 8312907398\\ \hline
%         $r_4$ & Lucy Kevin & 200 High St & 8314008741\\ \hline
%         $r_5$ & L. Kevin & 200 High St & 8004001341\\ \hline
%         $r_6$ & Lucy Kevin & 200 HIGH ST & 8004001341\\ \hline
%         $r_7$ & Smith Joh & 100 Mission St & 8004311928\\ 
%     \end{tabular}
%     \end{minipage}
%     }
    \subfigure[]{
    \begin{minipage}[h]{0.42\textwidth}
    \centering
    \includegraphics[width=0.9\textwidth]{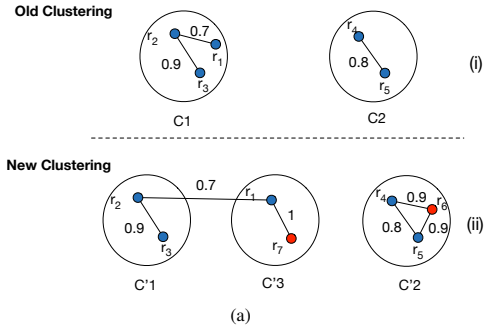}
    \end{minipage}
    }
\caption{A Motivating Example}
\label{fig:motivation}
\end{figure}

In this section, we present an example of using clustering algorithms and our dynamic clustering method to provide a better intuition for the rest of the paper.

The example setup is shown in Figure~\ref{fig:motivation}.
%
% Figure~\ref{fig:motivation}(a) shows the objects in the database (assume that there are 5 objects $r_1-r_5$ initially, and objects $r_6$ and $r_7$ are added later in the example.) 
Assume that there are 5 objects $r_1-r_5$ initially, and objects $r_6$ and $r_7$ are added later in the example.

Clustering can be used to group objects based on their similarity. The ``Old Clustering'' in Figure~\ref{fig:motivation}, for example, is the grouping for the initial five objects based on their similarity. The edges between the objects represent the similarity score, and the absence of an edge between two objects represents non-similarity. The clustering groups $r_1$--$r_3$ together in $C1$ and groups $r_4$ and $r_5$ together in $C2$. 
%This represents that three objects are matched (as part of C1) and two objects are matched (as part of C2). 

In the dynamic scenario, objects may be added, modified, or deleted. Consider that after the ``Old Clustering'' of the first five objects, two objects, $r_6$ and $r_7$, are added. The clustering of objects might change in reaction to the insertion of these two objects. One option to do this re-clustering is to perform a batch clustering algorithm that would compute the groupings from scratch for all objects. This, however, incurs significant overhead, especially for high-velocity dynamic workloads were data operations are continuous. An alternative is to incrementally assign objects to the clusters that are most similar to them---but without changing the structure of clusters. By computing the similarity, we may assign $r_7$ to C1 and $r_6$ to $C2$. However, such a method may not lead to an optimal solution. For example, it is better to split $\{r_1, r_2, r_3, r_4\}$ into $C'1$ and $C'3$ as illustrated in the lower part of Figure~\ref{fig:motivation} named ``New Clusterin''. Existing methods may overcome such inaccurate incremental clustering decisions, however, they require specific knowledge about the properties of the underlying batch algorithm. 

DynamicC proposes a dynamic machine learning-based method to allow both fast and accurate clustering decisions while not restricting the design to specific batch algorithms. Specifically, DynamicC would start by running an underlying batch algorithm that makes the clustering decisions during a training phase while observing its clustering decisions. After the training phase, DynamicC have built a model that given the current clustering and the incoming operations would predict the new clustering changes. In this example, assuming that DynamicC is already trained by the time $r_6$ and $r_7$ arrive; DynamicC would take as input global similarity metrics of the current clustering---of $C1$ and $C2$---and the newly arrived objects, $r_6$ and $r_7$. DynamicC would then make a prediction---using its model---of the changes that would apply to the clusters. For example, DynamicC starts by predicting that $r_7$ should merge with $C1$ and $r_6$ should merge with $C2$. Then, DynamicC predicts that $C1$ will split into $C'1$ and $C'3$. After this prediction, our heuristics algorithm makes the decisions on which objects from the original $C1$ would go to $C'1$ and which objects would go to C'3. In this scenario, as shown in Figure~\ref{fig:motivation} under ``New Clustering'', $r_2$ and $r_3$ are in $C'1$ and $r_1$ and $r_7$ are in $C'3$.

A nice feature of DynamicC's heuristics is that it can check whether a prediction to merge or split would lead to an improvement in clustering accuracy. This is done by computing the underlying batching algorithm's objective function for the predicted change. If the change yields an improvement in the objective function, then the change is made, otherwise, it is skipped. For example, consider the case in Figure~\ref{fig:motivation} where $C1$ is predicted to split. The heuristic algorithm checks whether the candidate split clusters would yield a better score for the objective function before applying the prediction.

In the rest of this paper, we provide the details of how DynamicC trains its ML model based on a batching algorithm and how the model is then used to make predictions and clustering decisions.

\subsection{Related Work}
%Dynamic structural clustering on graph \cite{ruan2021dynamic} try to evaluate their method in an approximate way.

Clustering (see~\cite{berkhin2006survey, rai2010survey} for a survey) is an unsupervised learning method which groups the objects based on  some similarity inherent among them into clusters. Traditional clustering techniques are mainly categorized as partitioning \cite{macqueen1967some,kaufman1990partitioning,awadelkarim2020prioritized}, density-based \cite{ester1996density,ankerst1999optics,rodriguez2014clustering}, and hierarchical clustering \cite{johnson1967hierarchical}. Note that this categorization of clustering is neither straightforward, nor canonical, and these categorization groups overlap in practice.

\textbf{Partitioning Clustering.} Partitioning clustering generally constructs a set of $k$ partitions, where each partition represents a cluster which
may be represented by a centroid or a cluster representative. Two renowned partitioning clustering are $k$-means \cite{macqueen1967some} and $k$-medoids \cite{kaufman1990partitioning}. These two clustering methods differ in how to represent the value of the objects in the cluster. 

\textbf{Density-based Clustering.}
The idea of density-based methods is to continue
growing as long as the number of objects in the neighborhood
exceeds some thresholds. Popular density-based clustering methods include DBSCAN \cite{ester1996density}, OPTICS \cite{ankerst1999optics}, DP \cite{rodriguez2014clustering} and so on.

\textbf{Hierarchical Clustering.}
Hierarchical clustering \cite{johnson1967hierarchical} is a method of cluster analysis which
works by building a hierarchy of clusters. Hierarchical clustering methods fall into two types: agglomerative and divisive. While agglomerative clustering proceed by series of merge operations on all the objects into groups until certain conditions are satisfied, divisive methods separate the objects successively into
finer (smaller) groups.

\textbf{Grid-based Clustering.} Grid-based clustering quantizes the object space into some segments (also cube, cell, or region) and then do space partitioning based on these segments.
A typical example of the grid-based approach is STING \cite{wang1997sting}.
This category is eclectic: it contains both partitioning and hierarchical
algorithms.

\textbf{Objective-based Clustering.} Objective-based clustering \cite{awadelkarim2020prioritized, bansal2004correlation, gruenheid2014incremental, macqueen1967some} finds clusters that minimize or maximize an objective
function. Typical objective functions in clustering formalize the goal of attaining high intra-cluster similarity and low inter-cluster similarity. The popular objective function based clustering includes $k$-means~\cite{macqueen1967some}, balanced $k$-way partitioning clustering~\cite{awadelkarim2020prioritized}, correlation  clustering~\cite{bansal2004correlation}, DB-index clustering~\cite{gruenheid2014incremental}, and so on. There are a lot of variants of the objective function based clustering as the objective function often depends on specific applications or problems. However, finding the optimal solution for such clustering is often NP-hard or intractable. This makes it hard to find a good dynamic algorithm for such clustering problems. 
%In addition, the objective function based clustering can be any of other clustering methods as it overlaps with many other clustering categories.

\textbf{Incremental Clustering and Dynamic Clustering.} 
Incremental clustering is less studied than the general batching approach. In incremental clustering, the focus is on performing clustering incrementally as new objects are added to the database. This is similar to our goal in dynamic clustering where we aim to perform clustering in response to changes (updates, inserts, and deletes) in the database. 

Existing dynamic approaches only work for specific clustering methods and may have high time complexity. For instance, Gan et. al.~\cite{gan2017dynamic} and Ester et. al. \cite{DBLP:conf/vldb/EsterKSWX98} propose some dynamic algorithms which are customized to the specific density-based method DBSCAN~\cite{ester1996density}. 
Some work (e.g. \cite{barbosa2016novel,nentwig2016holistic,sartorio2020dynamic}) focuses on designing new dynamic algorithms that efficiently deal with different tasks. Nevertheless, these works actually belong to ``batch algorithms'' in our scenario since they do not rely on other existing batch clustering algorithms.
Most similar to our work is the proposal by Gruenheid et. al.~\cite{gruenheid2014incremental}. That work proposes incremental approaches for re-clustering that are light-weight in comparison to an underlying batching method. However, for their approach to work, they require special properties on the objective function of the underlying clustering algorithm, such as locality and monotonicity. This restricts the applicability of these incremental methods to objective functions satisfying these requirements and excludes widely-used clustering approaches such as k-means, balanced k-way partitioning and DB-index---that does not satisfy the locality, monotonicity, exchangeability, and separability properties.

Our work, DynamicC, does not make stringent assumptions about the properties of the underlying clustering algorithms. This makes it applicable to a wide-range of clustering solutions unlike existing incremental methods. 
% In the evaluation section, we compare DynamicC with the methods proposed in~\cite{gruenheid2014incremental}.
%
Some clustering methods are naturally incremental (e.g. hierarchical clustering and grid-based clustering). That is, the incremental or dynamic algorithms for these clustering methods can be straightforward since new updates will not affect previous clustering result. For this reason, we do not evaluate these kinds of clustering methods in our evaluation section and focus on problems that are more challenging to make dynamic.

\section{Dynamic Clustering}
In this section, we begin by formally defining the dynamic clustering problem. Then, we introduce some background about the underlying batching-based clustering problems that will be used in the discussion and evaluation of our work.
\vspace{-5pt}
\subsection{Problem Definition}
\label{sec:ProblemDef}

\textbf{Clustering Problem.}
Given a set of objects $\textbf {D}$, clustering aims to group data $\textbf {D}$ into different clusters, where objects in a cluster are similar to each other. 

\textbf{Operations.}
We define three kinds of possible operations on data and discuss how they lead to triggering dynamic clustering: 
\begin{itemize}
    \item {\bf Adding:} When a new object is added, it may lead to one of the following re-clustering events: the object may join an existing cluster, it may be in a cluster by itself, or it may lead to splitting an existing cluster and then joining one of the resulting split clusters.
     
    \item {\bf Removing:} Removing a data object changes the similarity relations between objects in the corresponding cluster. This might lead to splitting or merging with another cluster. 

    \item {\bf Updating:} When an object is updated, this can potentially change the similarity scores between it and other objects. An update has a similar effect to removing data and adding new data which may lead to merging or splitting corresponding clusters.
    
\end{itemize}

We now formally define the problem of dynamic clustering:

\begin{definition} 
[Dynamic~ Clustering] Given the dataset $\bf {D}$ and its clustering results $\mathcal{L}_{\bf{D}}$, let $\bf {D_{new}}$ be a new set of objects with some updates on $\bf D$ where $\bf{D} \cap \bf {D_{new}} \neq \emptyset$. Dynamic clustering computes the clustering results of $\bf {D_{new}}$ based on $\mathcal{L}_{\bf{D}}$. The dynamic clustering is denoted by $F$ and  its clustering result is denoted as $F(\bf{D}, \bf{D}_{new}, \mathcal{L}_{\bf{D}})$.
\end{definition}

The goal of dynamic clustering is to find an efficient dynamic algorithm $\textit {F}$ such that we can get the same or similar results compared to that of the batch algorithm. Since an objective function is defined to evaluate the clustering results in this work, the goal of dynamic clustering is to make
$F(\bf{D}, \bf{D}_{new}, \mathcal{L}_{\bf{D}})$
$\approx B(\bf{D}_{new})$, where $B(\cdot)$ returns the clustering result of the batch algorithm. 
Dynamic clustering becomes challenging when new updates lead to changes in the clustering structure (clusters merging and/or splitting). Otherwise, the dynamic solution is straightforward as it only involves adding the new data to a cluster.

% \begin{table}[h]
% \scriptsize
% \centering
% \begin{tabular}{|c|l|}
% \hline
% \bfseries NOTATION & \bfseries   DESCRIPTION\\
% \hline
% $r\in R$  & A data$r$ in the dataset $R$\\\hline
% $r\in C$  & A data$r$ in the cluster $C$\\\hline
% $S_{intra}(C)$  & Intra-clustering similarity of cluster $C$\\\hline
% $S_{inter}(C, C')$  & Inter-clustering similarity between cluster $C$ and $C'$\\\hline

% \end{tabular}
% \caption{Notations used in this paper}
% \label{table:notion}
% \end{table}

\subsection{Preliminary}
\label{clustering:background}
Our work in this paper is applicable to any batch clustering algorithm. This is because we do not make assumptions about the objective function and build the machine learning model in DynamicC by observing (and learning) the cluster evolution. 

To better illustrate our method, we show how it is integrated with clustering methods that are based on optimizing an objective function (we call these methods \emph{objective-based}). But, we also evaluate our method on other kinds of clustering approaches in the experiment section to show its effectiveness. We focus on objective-based clustering for three reasons. First, these clustering methods (e.g. k-means \cite{macqueen1967some} and correlation clustering \cite{bansal2004correlation}) are widely used in various applications. Second, there are a lot of variants for objective-based clustering. This means that generating a dynamic method is needed due to the infeasibility of designing specialized methods for all variants. Third, the batch algorithm of this kind of clustering is often intractable (NP hard or complete) or has high complexity, therefore, a light-weight method for dynamic clustering would offer a significant performance advantage.

In the following, we briefly describe the clustering properties for objective-based clustering that we use in our formulations.
%\footnote{Finding an appropriate objective function for optimization can be done with some ML models. We do not focus on how to find the best objective functions as it is orthogonal to our work.} %// Faisal: yes, good point. But I think that now that we clarify this early on in the related work section, we do not need to repeat it here.

% Each cluster in \textit{Correlation Clustering} and \textit{DB-index clustering} is associated with two types of measures: \emph{intra-cluster} similarity and \emph{inter-cluster} similarity. While intra-cluster similarity measures the similarity of objects in the same cluster, inter-cluster similarity measures the similarity of objects across clusters. Both these clustering methods aim to maximize the intra-cluster similarity while minimizing the inter-cluster similarity.

\begin{enumerate}[1)]
\item    \textbf{Intra-cluster similarity:} For each cluster $C$, the intra-cluster similarity is the sum of the similarity between objects in the same cluster: $S_{intra}(C)=  \sum_{r, r' \in C}  sim(r, r')$, where $r$ and $r'$ are any two different objects in the cluster $C$.
%complement of   1-
\item  \textbf{Inter-cluster similarity:} For each pair of distinct clusters $C$ and $C'$, the inter-cluster similarity is the sum of the similarity between objects across the clusters: $S_{inter}(C, C')= \sum_{r \in C, r' \in C', C'\neq C } sim(r, r')$. 
% \Change{$S_{inter}(C)= \frac { \sum_{r \in C, r' \in C', C'\neq C } sim(r, r')} {2}$}
\end{enumerate}

\textbf{Objecive Function.}
To better explain our method with examples, we list the objective function of correlation clustering below. 
The objective function of correlation clustering---which we will use in some of our example scenarios---is below:

\begin{equation}
\label{eq:correlation}
\begin{aligned}
F(\mathcal{L}_{R}) =& \sum_{C\in C_R}{(1- S_{intra}(C))}+ \\
&\sum_{C\in C_R, C'\in C_R, C\neq C'}{S_{inter}(C, C')}
\end{aligned}
\end{equation} 
Correlation clustering aims to minimize the complement of intra-similarity and the inter-similarity. 

Although different clustering methods use different metrics (e.g. mean) to evaluate their objective function, we focus on the general idea of clustering that $(1)$ objects in the same cluster are similar to each other with intra-similarity, and $(2)$ objects in different clusters are not similar to each other with inter-similarity.

\begin{figure}
\setlength{\abovecaptionskip}{0.cm}
\setlength{\belowcaptionskip}{-0.cm}
\centerline{\psfig{figure=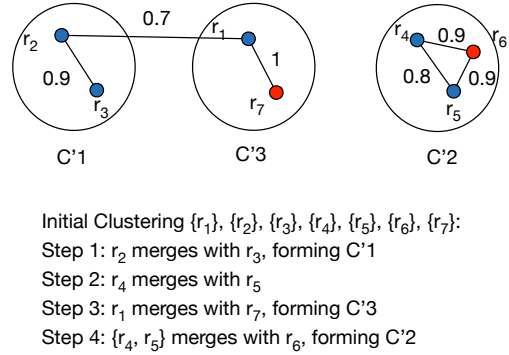, width=0.4\textwidth} }
\caption{Possible clustering of seven objects from scratch.}
\label{fig:new}
\end{figure}

% \begin{figure}
% \centerline{\psfig{figure=pics/initialcluster.eps, width=0.35\textwidth} }
% \caption{Possible clustering of clustering objects $r_1$ to $r_5$ using a batch algorithm}
% \label{fig:initial}
% \end{figure}

\section{Monitoring Cluster Evolution}
\label{sec:historical}
In this section, we present our method to represent and monitor cluster evolution relations while running the underlying batching clustering algorithm. This step is important as cluster evolution information is going to be used to train our ML model that will predict future clustering changes as we show in Section~\ref{sec:model}. We begin by an overview of cluster evolution operations. Then, we introduce how to extract and represent the cluster evolution information from one round of batch execution from scratch. After that, we present how to represent cluster evolution across batch clustering rounds.

% After that, we present how to manage those clustering changes in order to connect the clustering results from one snapshot to its next snapshot with the updates. 

\subsection{Cluster Evolution Operations}
We define two types of evolution operations: \textit{merge} evolution and \textit{split} evolution.
The \textit{merge} evolution means that two clusters merge into one cluster.
The \textit{split} evolution means that a cluster is split into two clusters.
One evolution can involve more than two clusters. For example, more than two clusters may merge into a cluster when executing a batch algorithm. 
In such a situation, a change can be divided into multiple merge evolution operations that only involve two clusters. As an example, suppose $r_1$, $r_2$ and $r_3$ are merged together into cluster $C$. We can represent this merge of three clusters as merging $r_1$ and $r_2$ into $C'$ and then merging $r_3$ with $C'$. Thus, using an abstraction of a merge/split operation involving two clusters only is sufficient to represent merges and splits involving more than two clusters.

There is another type of evolution called \textit{move} evolution. The \textit{move} evolution means that some objects move from one cluster to another cluster. The \textit{move} evolution is equivalent to the combination of the \textit{split} and \textit{merge} evolution. For example, moving $r_1$ from $C1$ to $C2$ in the ``old clustering'' in Figure~\ref{fig:motivation} is equivalent to splitting $C1$ into $\{r_1\}$ and $\{r_2, r_3\}$ first and then merging $\{r_1\}$ with $C2$.  Therefore, \textit{merge} and \textit{split} evolution operations are sufficient to express all the possible changes. There is a special situation that splitting a cluster does not improve the score but the move operation does. However, we seldom find such instances which leads to such changes being insignificant for model training. 

%Also, we introduce how to deal with such special situations in our approach in Section\ref{sec:algorithm}.

% The difference between the two kinds of changes is that the second method generates the changes at the coarse granularity, while the first one generates changes at the finer granularity. The second method is more accurate as it check whether two clusters should be really merged or split. But one problem is that we may not have much training data to train a good ML model because the number of changes can be quite small.  The first method can not guarantee that so it may need to iteratively check the clusters leading to much cost.   

\subsection{Cluster Evolution From Scratch}
Here, we begin by showing how cluster evolution can be represented when running a batch algorithm from scratch (the evolution starts from an initial clustering where every object is in a cluster by itself.)
Reaching the final clustering result is often a step-by-step process. As illustrated in Figure~\ref{fig:new}, a batch algorithm may take 4 steps to cluster seven objects from scratch towards the final clustering $\{C'1, C'2, C'3\}$. Each step moves towards a better clustering result. 
%This step-by-step process depends on the used batch algorithm.
To judge whether a clustering outcome is successful, we compute a score by an objective function.
% such as the one in Equation~\ref{eq:correlation}. 
When a batch clustering algorithm terminates, it means that it reached an optimal solution.

\begin{example}
Consider the objects in Figure~\ref{fig:new} and apply correlation clustering to them. As described in Section~\ref{clustering:background}, the goal is to minimize an objective function $F$. The similarity among objects is also shown in Figure~\ref{fig:new}. Having no edge between two objects represents a similarity score of $0$. Initially, we have the initial clustering $\mathcal{L}_{1}=\{\{r_1\},\{r_2\},\{r_3\},\{r_4\},\{r_5\},\{r_6\},\{r_7\}\}$ (each object is in its own cluster) and $F(\mathcal{L}_{1})=0.9*3+0.8+0.7+1=5.2$. Suppose $r_1$ merges with $r_7$ forming $C'3$, we now have the clustering $\mathcal{L}_{2}=\{C'3,\{r_2\},\{r_3\},\{r_4\},\{r_5\},\{r_6\}\}$. Since $F(\mathcal{L}_{2})=0.9*3+0.8+0.7+(1-1)=4.2<5.2=F(\mathcal{L}_{1})$, it indicates that $\mathcal{L}_{2}$ is a better clustering than $\mathcal{L}_{1}$. This process continues until the clustering reaches the optimal clustering. 
\end{example}

\subsection{Cluster Evolution Across Rounds}
As objects are added, removed, and updated, the clustering of objects might become stale. To overcome this, a new round of the clustering algorithm should be performed. A batch clustering algorithm would process every round by clustering from scratch. However, we want to come up with a representation of cluster changes that captures the differences between the cluster results of the current and new clustering. The reason for this is that we want to train our machine learning model to mimic these changes from one round to the next round without having to re-cluster from scratch.

For instance, consider we are monitoring the evolution of a batch clustering algorithm that went through two rounds. In the initial round, the batch algorithm clustered objects $r_1$ to $r_5$ and resulted in the clustering shown in Figure~\ref{fig:motivation}. This initial clustering can be represented using the method we presented above for clustering from scratch. However, when objects $r_6$ and $r_7$ are added, the batch algorithm clusters all the objects again from scratch to arrive at the clustering in Figure~\ref{fig:new}. The challenge here is that we do not want to capture the steps in this second round of clustering from scratch---instead, we want to capture the steps of the difference between the clustering from Figure~\ref{fig:motivation} to the clustering in Figure~\ref{fig:new}.

% Some batch algorithms (e.g. hill-climbing algorithm) start from an initial clustering which might be far crude than clustering of in the previous round. 

%In addition, the model can be updated instantly as new changes come in the latter rounds.

% \subsection{Changing Process Simulation}
% When we run a batch algorithm from scratch, the changes towards the final clustering results happen from an primitive state rather than the clustering results of its previous snapshot. For instance, as illustrated in Figure~\ref{fig:example2}, we first take each node as a single cluster and then merge these single clusters to larger clusters. However, these possible changes shown in Figure ~\ref{fig:example2} seem unhelpful because we can not see explicit changes happening from previous clustering results (in Figure~\ref{fig:example1}) to current clustering results. As defined in Section\ref{sec:ProblemDef} (Definition 3.1), the dynamic method relies on and starts with previous clustering results. To this end, we first need to manage these changes, making them helpful to connect previous clustering results to current results. 

% \begin{example}
% Consider the example in Figure\ref{fig:example2}, suppose we run a batch algorithm from scratch and then we get the following cluster changing history.
% \begin{enumerate}
%     \item [Step 1:] $r_1$ merges with $r_7$, forming C3
%     \item [Step 2:] $r_2$ merges with $r_3$, forming C2
%     \item [Step 3:] $r_4$ merges with $r_5$
%     \item [Step 4:] [$r_4, r_5$] merges with $r_6$, forming C2
% \end{enumerate}
% \end{example}

To represent the cluster evolution of the changes only from one clustering to the next, we come up with an algorithm that generates a series of clustering steps (merges and splits) that would transform the old clustering to the new one. The goal of this algorithm is to find a small number of steps to represent this evolution from a clustering to the next efficiently. The algorithm proceeds in the following two phases:

Phase 1: We look for the changes that are relevant to the objects that were added, modified, or removed in this round (\emph{e.g.}, $r_6$ and $r_7$ in our example). We take that change and add it to the list (called \emph{transformation list}, of steps to transform from the old clustering to the new one.
%(we will call this list the \emph{transformation list} hereafter)
If an object has undergone multiple changes during a round, then we only add the latest one to the transformation list. 
% For example, suppose that $r_6$ first merges with $r4$ and then $\{r_6, r_4\}$ merges with $r_5$. There would be two changes about $r_6$. However, the first one is unnecessary since $r_4$ and $r_5$ may be already merged in the old clustering. 

Phase 2: The algorithm checks the rest of the changes that pertain to the remaining (old) objects (objects that were not added, updated, or deleted in this round.) For each cluster $c$ in a change (a merge change contains two clusters and a split change only contains one original cluster and two new clusters), if $c$ contains one or more old objects and it does not exist in the old clustering, we then split the $n$ clusters which have overlapping objects with $c$ in the old clustering. A cluster $c'$ in the old clustering is split into $c'\cap c$ and $c'-(c' \cap c)$. If $c'\subset c$, $c'$ is split into $c'$ and $\emptyset$. Therefore, we do not really split $c'$ in this case. Finally, we merge these intersection clusters one by one, forming the cluster $c$ and get $n-1$ merge changes.

% \begin{algorithm}
%       %\SetAlgoNoLine 
%       \LinesNumbered
%       \SetKwInOut{Input}{\textbf{Input}}\SetKwInOut{Output}{\textbf{Output}}
      
%       \Input{
%           Previous clustering results $Cl_{old} = \{C_1, C_2, ..., C_m\}$; \\
%           Relevant Updates $R_{update} = \{r_1, r2, .., r_n\}$; \\
%           Cluster evolution with new updates by using a 
%           batch algorithm: $CE_{batch} = \{Ch_1, Ch_2, ..., Ch_p\}$.}
%     \Output{
%         Cluster Evolution: $CE_{new} = \{Ch_1, Ch_2, ..., Ch_q\}$.}
%     \BlankLine
%     \For {$Ch\in CE_{batch}$}{
%     \If{$Ch$ does not contain any data$r\in R_{update}$}{
%     remove $Ch$ from $Ch_{batch}$}
    
%     }
%     \For {$Ch \in CE_{batch}$}{
%     \For{each cluster $c$ in $C$}{
%     \For{$C\in Cl_{old}$}{
%     \If{$c\cap C = s \neq \emptyset$}
%     {\If{$s\subset C$}{
%     split $C$ into $s$ and another cluster $C-s$ \;
%     add $s$ and $C-s$ into $Cl_{old}$ and remove C from $Cl_{old}$\;}
%     }
%     }
%     }
%     }
    
%   \Return {$Ch_{new}$} \;

%     \caption{Cluster Evolution Transformation Algorithm}
% \label{algo:evolution}
% \end{algorithm}

The following is an example of how we derive these clustering steps across rounds.

% \Change{(Need to revise)

% In order to check if a change $\Delta(C)$ is valid, we first fix other clusters that are irrelevant to $\Delta(C)$. If the score becomes larger with this change, then we say the change is valid because it leads to a better clustering result.}

\begin{example}
\label{ex:2}
Consider an example of trying to come up with clustering steps to represent the evolution from the clustering in Figure~\ref{fig:motivation} to the clustering in Figure~\ref{fig:new}. 

\textit {Phase 1:} Since only steps $3$ and $4$ are relevant to new objects $r_6$ and $r_7$, we keep them and get rid of steps $1$ and $2$. Thus, we have the following two changes initially.

{\bf Change 1:} $r_1$ merges with $r_7$, forming $C'3$.

{\bf Change 2:}$\{r_4, r_5\}$ merges with $r_6$, forming $C'2$.

\textit {Phase 2:} Change 1 contains two clusters $r_1$ and $r_7$ and Change 2
contains two clusters $\{r_4, r_5\}$ and $r_6$.  Since $r_6$ and $r_7$ are not old objects, we do not need to consider them. Because $r_1$ does not exist exactly in the old clustering $\{C1, C2\}$, we then split C1 into $r_1$ and $\{r_2, r_3\}$ as $C1 \cap r_1 = r_1$ and $C1-r_1 = \{r_2, r_3\}$. Therefore, we have the following change.

{\bf Change 3:} Split C1 into $r_1$ and $\{r_2, r_3\}$.

Since $\{r_4$, $r_5\}$ is the same as C2, i.e. $\{r_4$, $r_5\}$ exists exactly in $\{C1, C2\}$, we do not need to generate another change.
\end{example}

% \Binbin{(Faisal: Let's talk about this paragraph below when we meet. I am not sure I completely understand it.)}
Note that our algorithm derives the steps to make the transformation from the old to the new clustering but it does not necessarily come up with the right ordering for these steps. This is because such ordering is not necessary for training the machine learning model that trains by observing the changes independently (as we observe in Section~\ref{sec:algorithm}). In addition, as can be seen in the next section, deriving the order of cluster evolution is intractable or does not help improve the efficiency. Therefore, we adopt a simpler solution to derive the steps of the transformation across rounds.

\section{DynamicC ML Model}
\label{sec:model}
In this section, we present the details of training our machine learning model using cluster evolution history information. Then, we discuss the effectiveness of machine learning models in our scenario. Lastly, we explain how to guarantee that the machine learning model is accurate.
We emphasize that our goal is not to design new ML-based batch clustering models or new feature selection methods. Rather, our goal is to propose a light-weight machine learning model to be augmented with an existing batch clustering algorithm to expedite the process of making clustering decision.

\subsection{DynamicC Feature Selection}
Here, we present the features we use in DynamicC's machine learning model. These features are independent from the underlying batch clustering algorithm and depend only on the observed characteristics of the resulting clustering of objects. We picked these features to describe global characteristics of the clustering to capture the current state without using too much information. The following are these features:

\noindent {\bf Intra similarity.} This feature represents the average similarity between attributes in the same cluster. This measure gives a score of the \textit {cohesion} of a cluster. Therefore, it is a good feature to predict whether a cluster would split or merge.

\noindent {\bf Maximal inter similarity.} This feature represents the maximal similarity among all the average similarity scores between a cluster and the other different clusters. This is a measure of correlation between different clusters. This feature mainly facilitates predicting whether the similarity between clusters would trigger a merge.

\noindent {\bf Cluster size.} This feature represents the size of a cluster. This measure can transform the average similarity of a cluster into the sum of the similarity of a cluster by multiplying by the cluster size. 

% These features can be easily selected as important ones without much overhead in feature engineering for ML models. As we emphasize in the beginning of this section, we show how the ML model help us do dynamic datalinkage even with naive feature selection techniques.

\subsection{Training the Model}
\label{subsec:training}
The training phase of DynamicC aims to build a machine learning model to enable predicting future cluster changes. This training is performed on a dataset of historical cluster evolution changes that are collected by running a batch clustering algorithm on the desired workload---using the methods in Section~\ref{sec:historical}.
All evolution steps are represented via {\bf merge} and {\bf split} operations that involve two clusters only. As we have shown in Section~\ref{sec:historical}, these two operations are sufficient to represent all cluster evolutions.

\begin{sloppypar}
An important question in designing and training the machine learning model is how to represent data points in terms of the input to the model and the decisions to be made. A straight-forward representation given our problem is to train the machine learning model to predict whether every pair of clusters should be merged (or split). The problem with this representation is that the number of pairs is high---$n*(n-1)/2$, where $n$ is the number of clusters. This will produce an intractable machine learning model that is difficult to train as the state space to be learned is large.
\end{sloppypar}

Instead, to make the training the machine learning model tractable, we train the model to predict whether a cluster $C$ should be merged or split only. The model needs not predict which clusters should be merged with $C$ or how a cluster $C$ should split (later, we present algorithms to make these decisions given the machine learning outcome and introduce how to guarantee the quality of these solutions in Section~\ref{sec:guarantee} and Section~\ref{sec:algorithm}).

Formally, for the \textit{Merge} model, each value in the training data is a $5$-dimensional vector $(d_1, d_2, d_3, d_4, d_5)$ about a cluster $C$:
\begin{enumerate}[1)]
    \item $d_1$ is the average intra-similarity of $C$ and $d_1\in [0, 1]$
    \item $d_2$ is the maximal inter-similarity of $C$ and $d_2\in [0, 1]$. 
    \item $d_3$ is the size of $C$ and $d_3\in [1, +\infty)$. 
    \item $d_4$ is the size of the cluster $C'$, where the average similarity between $C$ and $C'$ is maximal for $C$ and $d_4\in [1, +\infty)$. 
    \item $d_5$ is the output label which indicates if $C$ should be merged or split and $d_5\in \{0, 1\}$ where $0$ indicates that $C$ should not be merged and $1$ indicates that $C$ should be merged. 
\end{enumerate}

For the $\textit {Split}$ model, each value in the training data is a $4$-dimensional vector $(d_1, d_2, d_3, d_5)$ about a cluster $C$ (what each entry in the vector represents matches the definitions above for \textit{Merge}). The input of $\textit {Split}$ does not have $d_4$ because split evolution refers to one cluster. 

% \begin{example}
% Continuing with example~\ref{ex:2}, for the cluster evolution Change $1$, we can get two $5$-dimensional vectors $(0, 0.9, 1, 2, 1)$ and $(0.8, 0.9, 2, 1, 1)$ related to the two clusters $\{r_6\}$ and $C_2$. For the cluster evolution  Change $2$, we can get one $4$-dimensional vectors $(0.53, 1, 3, 1)$ 
% \end{example}

The model is trained separately based on merge evolution and split evolution. That is, the training data of the merge model only contains the merge evolution and the training data of the split model only contains the split evolution. 
% This is because the dimensions of the input for merge and split models are different which prevents training them together.
%For effective training, we uniformly sample from all negative instances.

\begin{figure}
\setlength{\abovecaptionskip}{0.cm}
\setlength{\belowcaptionskip}{-0.cm}
\centerline{\psfig{figure=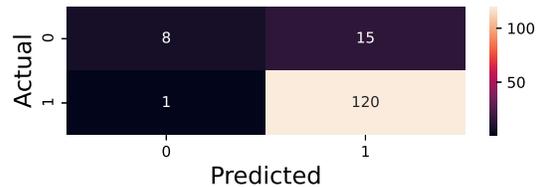, width=0.48\textwidth} }
\caption{Heatmap of prediction performance on the testing data}
\label{fig:heatmap}
\end{figure}

\subsection{Negative Sampling}
Instances of cluster evolution are defined as positive samples where the label is set to $1$ for training in our case. In order to learn an effective ML model, we need negative samples for training. The negative samples are those clusters which are unchanged when the batch algorithm is performed. Compared with the amount of positive samples, the number of negative samples is often much larger. But we care more about those clusters involved in the connected components since these clusters are checked more times during the execution of a batch algorithm. We call these clusters ``active'' clusters.
Our strategy is to uniformly sample from all negative instances but we give higher weight to those ``active'' clusters (the weight is set to $0.7$ for active clusters and set to $0.3$ for non-active clusters in our experiments). We set the number of negative samples to be equal to that of the positive samples for effective training. In addition,  we remove those old samples when the size of training data become too large. This is because the old samples become less and less important in the dynamic scenario and reducing the size of training data could keep the ML model efficient and effective.

% \subsection{Effectiveness of ML models}
% \label{sec:effective}
% \Binbin{(Faisal: Let's talk about this subsection tomorrow to clarify what it means.)}
% Most ML models assume new data follows existing patterns. This assumption may limit the opportunity for ad-hoc clustering patterns. However, in dynamic datalinkage, this problem can be greatly alleviated.
% %
% Firstly, the volume of changed objects is typically much smaller than that of current objects. Otherwise, running a batch algorithm over the whole data would be more efficient because the overhead of running a batch algorithm on the changed objects will be approximate to that of running a batch algorithm over the whole data. When the number of the updated objects is much less than that of the unchanged objects, the updates will not cause significant change in existing patterns. 
% %
% Secondly, the relationship between any two objects depends directly or implicitly on their similarity (e.g. intra and inter similarity) which is stable throughout clustering changes. This prevents outliers from emerging suddenly without any prior signal.
% %
% To some extent, these two features guarantee that the patterns of the clustering will not change dramatically. 
% %In this paper, we use the ML model to predict whether or not a cluster will change.

\subsection{The Objective of the ML Model}
\label{sec:guarantee}
In this section, we discuss the objective when training DynamicC's ML model. Typical ML models train to optimize accuracy. However, we observe that training for recall leads to better results for our problem. In the rest of this section, we discuss this observation and its implications to the training of the ML model.

First, we present the \emph{quality measures} that are relevant to this discussion: accuracy, precision and recall.

\noindent{\bf Accuracy and Precision. } While accuracy means the ratio of the number of correct predictions to the total number of input samples, precision represents how precise your model is out of those predicted positive. As an example, Figure~\ref{fig:heatmap} shows a heat map for model prediction of $144$ clusters. Label $0$ represents negative predictions and label $1$ represents positive predictions. Since $16$ ($1+15$) clusters are predicted wrongly and $128$ clusters are predicted correctly, the accuracy is $128/(16+128)=0.889$. $15$ negative clusters are predicted incorrectly, so the precision is $120/(120+15)=0.89$.

%However, the recall of the model is a better metric in our case. 

\noindent{\bf Recall.} Recall calculates how many of the true positives (or actual positives) the model captures correctly. So the recall in our scenario is $120/(1+120)=0.992$. 

Although we can not guarantee the quality of any of the three metrics of a ML model, these metrics inspire us to think differently of how to apply ML to our problem.
We only need to find all the clusters which are ought to change (and we are not affected by finding false positives as we can swiftly check whether these are false positives using the objective function, which we describe next). That is, we are trying to find all positive clusters. This means that recall is the most relevant metric for our problem and we use this observation to influence how we train the DynamicC ML model.
In the rest of this section, we present the details of how we avoid false positives and how this leads us to the observation that we ought to train DynamicC's ML model to maximize recall instead of accuracy.

%\Binbin{\noindent{\bf Accuracy? Performance Guarantee}}

\textbf{Avoiding False Positives.}
Incorrectly predicting the negative clusters as positive clusters can be amended with a swift verification after the prediction. Specifically, we can verify whether we should apply the change that is suggested by the ML model by computing whether the objective score will improve with this change; if it will not improve, then we do not apply that change and the incorrect prediction is avoided. For instance, in Figure\ref{fig:new} suppose the ML model predicts that $r_7$ should merge with $\{r_2, r_3\}$, we can assign $r_7$ to $\{r_2, r_3\}$ and compute an objective score for the new suggested clustering$\{\{r_7,r_2, r_3\}, \{r_1\}, C'2\}$ and compare it with $F(\{C'1, C'2,  C'3\})$.
Computing an objective function adds an overhead but guarantees the correctness of the decision of a merge change or split change. 
% Faisal (Feb 22): removed the following sentence -- I am not sure what it means or if it adds to the discussion. Return it and discuss with me to rephrase if you feel it is important.
%In addition, the model also benefits from such a strategy when the wrong samples are corrected by checking the objective function score. This is because the model can not adjust itself with new labels if all the new labels are predicted by the ML model itself.  

\textbf{Maximizing Recall Instead of Accuracy.}
Now, our ML model problem becomes about maximizing recall rather than maximizing overall accuracy. Actually, we can reach high recall if there are no other constraints. (As an extreme example, we can make a ML model predict all the samples as positive such that the recall is $100\%$. However, this strategy would lead to bad accuracy and performance due to having to verify and amend all incorrect predictions.) The strategy we adopt is that we train the ML model to have nearly $100\%$ recall while maximizing accuracy. Training such a model is not straight-forward with machine learning techniques. Next, we explore this question of training for both recall and accuracy.
%and degenerate the ML model. For instance, a ML model with accuracy less than $50\%$ would be invalid for a binary classification task.

% In addition, the ML model can predict for a cluster with its confidence. This kind of information can help us determine which clusters should be checked first in terms of efficiency.
% To summarize, the problem is how to train a model which has the highest accuracy and the $100\%$ recall meanwhile.

% When we train the ML models with open-source machine learning libraries, these tools  provide us with a generally good model (e.g. the one with the highest accuracy). Although, we can redefine a loss function to make the model reach $100\%$ recall on the validation dataset, this method may lead to an ineffective like the classifier3 in Figure~\ref{fig:classifier}. Also, we lose the opportunity to explore the trade-off between the quality and the efficiency of datalinkage as we do not know which are the candidate search space that we should check.

\noindent{\bf Trade-off Exploration.} 
We now study how to control the trade-off between the quality and efficiency of clustering.

Usually, a ML model predicts if an item $t$ belongs to a class $c$ with a probability $P(t=c)$.
% Even though two items $t$ and $t'$ may be predicted as $c$, the probabilities $P(t=c)$ and $P(t'=c)$ can be very different because the final prediction is also determined by a threshold.
The following equation illustrates how a ML model predicts a binary classification task.
\begin{equation}
\label{accuracy}
{Label(t)=}
\left\{\begin{array}{ll}
1 & \textrm{if $P(t=1) \geq \theta$}\\
0 & \textrm{otherwise}
\end{array} \right.
\end{equation}
where $t$ is the item to be predicted . In our case, an item $t$ is a cluster $C$, therefore we will use $P(C=c)$ instead of $P(t=c)$ in the rest of the paper.
%The probability indicates the confidence on the decision. 
% In Figure~\ref{fig:classifier}, suppose classifier $1$ is the ML model, then being away from the dotted line of classifier 1 would lead to a higher confidence---and thus probability---on the decision. 
%
Based on this mechanism, we can check which probability values those positive samples have on the training data and set $\theta$ as the minimal value of those positive samples. In this way, the probabilities of all positive samples are no less than $\theta$, leading to $100\%$ recall on the training data.

\begin{figure}
\setlength{\abovecaptionskip}{0.cm}
\setlength{\belowcaptionskip}{-0.cm}
\centerline{\psfig{figure=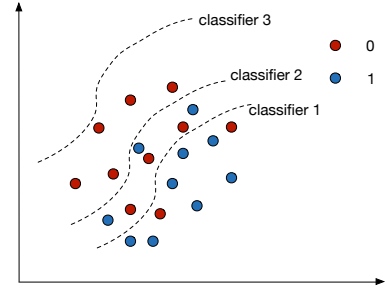, width=0.3\textwidth} }
\caption{Illustration of different classifiers in 2-D space}
\label{fig:classifier}
\end{figure}

The trade-off between the accuracy and efficiency can be done by setting a different $\theta$ value for prediction. The smaller the $\theta$ is, the more clusters we need to check. Moreover, the smaller $\theta$ often means higher recall and lower precision.
Note that we do not change $\theta$ value when training the ML model. We manipulate $\theta$ only when using the model to control the trade-off between accuracy and efficiency.
\begin{example}
As shown in Figure~\ref{fig:classifier}, classifier $1$ is the best model in terms of the accuracy, but three positive clusters (or nodes) are missed. Classifier $2$ is the best model in terms of recall. It finds all positive clusters and only needs to check $6$ more clusters than classifier $1$. The recall of classifier $3$ is also $100\%$, but it needs to check $11$ more clusters than classifier $1$.
\end{example}

\section{Dynamic Algorithm} 
\label{sec:algorithm}
In this section, we present the algorithm for dynamic clustering that uses the model we developed in Section~\ref{sec:model}. The algorithm handles three types of operations: add, remove, and update. When an operation is called, the corresponding object incurs initial processing to include it in a cluster. Then, the model runs and predicts whether a merge or split algorithm should be performed. We present the details of initial processing and the merge and split algorithms, followed by the \emph{full algorithm} that puts them all together. 

\subsection{Initial Processing}
\label{sec:initial}
%\noindent\textbf{Initial processing:}
When a data operation is performed, we first include the new object information within the clusters based on the type of operation:
\begin{itemize}
\item {\bf Adding new objects:} If new objects are added, then we consider each object as a new cluster with one object only. 

\item {\bf Removing an object:} This operation leads to removing the object from their current cluster. 

\item{\bf Updating an object:} When an object is updated, it is removed from its current cluster and placed in a new cluster with no other objects. This scenario can be represented as a remove-add sequence.
\end{itemize}
After the three types of operations are done, we perform the merge algorithm followed by the split algorithm as we describe in the rest of this section.

\subsection{Merge Algorithm}
%\noindent{\bf Merge algorithm:} 
The challenge with the merge algorithm is that the ML model only predicts what clusters are candidates for merge operations. However, the model---for tractability reasons we mentioned in Section \ref{subsec:training}---cannot directly predict which cluster it should merge with as the model is a binary classifier. Checking the pairwise clusters takes quadratic time (i.e $O(n^2)$) with the number of clusters. Thereby, pairwise checking can incur significant overhead when the number of candidate clusters has not been reduced significantly.

To reduce this overhead, we make an observation that if two clusters are ought to merge, it is likely that they are both predicted to be merged in our model. This observation leads to reducing the space of candidate clusters. That is, the clusters predicted as ``merge'' can only merge with those clusters which are also predicted as ``merge''. Therefore, the merge algorithm starts by constructing a set $Cl_{merge}$ that contains all clusters that are predicted to merge. 
% We put those clusters which are predicted as $1$ by the merge model into the candidate cluster set $Cl_{merge}$. 
%
Next, we determine which clusters in $Cl_{merge}$ should be merged together. To find this out, we start with a cluster $C$ in $Cl_{merge}$ and find the best candidate to merge with it. This can be done by checking whether merging with every other cluster in $Cl_{merge}$ would lead to a new stable cluster. Specifically, a stable cluster would be one that is not likely to merge again. This can be checked by observing $P(C_{new}=1)$, where $C_{new}$ is the result of merging $C$ and another cluster $C'$. The merge that minimizes $P(C_{new}=1)$ would be the best as it means that it results in the most stable clustering.
%first investigate what we can do to make the probability $P(C=1)$ decrease so that a cluster $C$ will be predicted as $0$ (meaning that it is predicted not to merge). If no clusters are predicted as $1$, it means the clustering is stable and we do not need to do any further merge operations. 
%
%Based on this observation, In our merge decisions, we focus on reducing the probability of being predicted as 1 for a cluster in order to make the cluster stable (or unchanged). For this purpose, we can find out the main factors that could significantly reduce such prediction probability. 
%
Checking the effect on $P(C_{new}=1)$ can be obtained from a ML model in a lightweight way through the coefficients of ML models. In our experiments, we find that the maximal inter similarity and the size of the clusters have respectively high weights in the learned ML model when we do merge predictions. 
% Inter similarity is expected as the important factor to judge if two clusters should be merged. However, the size of the clusters are often ignored when we make merge operation decisions.
%

In the following, we describe the merge algorithm.
% So, our strategy is that $C$ tries to merge with the cluster $C'$ which can maximally reduce the probability $P(C=1)$ by removing $C'$ from the existing clustering. 
% This is because when $C$ and $C'$ are merged, the parameters of $C$ will change since $C'$ can be seen as being removed from its candidate set. \Change{Need further explanation.}
%
Denote $Cl_{initial}$ as the initial clustering  and $Cl_{merge}$ as the cluster set in which clusters are predicted as $1$ by the merge model. The merge algorithm is performed as follows: 

\begin{itemize} 
    \item For each cluster $C\in Cl_{merge}$, find the cluster $C'\in Cl_{merge}$ which can maximally reduce the probability $P(C_{new}=1)$ by merging $C$ and $C'$. Evaluate if merging $C$ with $C'$ can generate a better clustering by checking the objective score. If so, remove $C$ and $C'$ from $Cl_{merge}$ and merge them in the new clustering. Otherwise, remove $C$ from $Cl_{merge}$.
\end{itemize}

Algorithm~\ref{algo:merge} shows how the merge algorithm works. First, the merge algorithm uses the \textit {merge} model to predict which clusters will be merged and put these clusters in the set $Cl_{merge}$ (line $2$). For each cluster $C\in Cl_{merge}$, it checks whether $C$ can merge with the selected cluster in $Cl_{merge}$ except itself (line $5$). If $C$ merges with a cluster $C'\in Cl_{merge}$, then $C$ and $C'$ are removed from $Cl_{merge}$ (line $6-8$). Otherwise, it removes $C$ from 
$Cl_{merge}$ (line $11$). This process continues until $Cl_{merge}$ is empty (line $3-13$).

After the merge algorithm terminates, we perform the split algorithm. We do merge algorithms first because each new object is taken as a new single cluster initially. It is more likely that these clusters can be merged with others rather than being split.

% The setting of threshold $k$ is to prevent that some clusters can never merge with other clusters successfully but we have to repeatedly check these clusters. 

% The following are some strategies. First, we tend to merge those clusters with small size, this is because small clusters will not changes the parameters of large clusters significantly. For example, suppose $r_1, r_2, r_3$ are in the cluster $C1$ and have an average similarity of $0.9$. Also assume $sim(r_4, r_5)=0.8$, $Avgsim(r_4, C1)=0.9$ and $Avgsim(r_5, C1)=0.7$. If $r_4$ can be successfully merged with $r_5$ since they have a respectively high similarity temporarily. However, $r_4$ and $r_5$ may be split if $C1$ merge with another cluster, forming a larger cluster. This strategy seems a good way to avoid the ``move" operation.

\subsection{Split Algorithm}
%\noindent{\bf Split algorithm:}
% Here, we only need to check $k$ times where $k$ is size of the cluster. Each time, we first select the one that unsimilar to most of them. Notice that we now have two clusters. Next, we choose the one similar to the single cluster. This is because we want to check if this cluster can be split into larger clusters. The strategy also shows the balance of the size of the clusters.
If cluster $C$ is predicted as 1 (i.e. ``split'') by the split model, we then check whether splitting it into two clusters would generate a better clustering. Before we introduce the detailed strategy, we first describe the possible scenarios when an existing cluster $C$ will split in response to a new data operations.
%
% \Binbin{the distinction and description is a bit unclear. Try to think of a concise way to say it that clarifies the difference. Also, for the second one, why didn't the datago to C' directly? I think one way to encompass these cases is to say that a datafirst joins the closest cluster. Then, the cluster becomes "too big", containing diverse objects, so it splits. The second case is not clear to me given the question I wrote above} 
%
When a new object arrives, it first joins the closest cluster. Then, the cluster becomes ``too big'', containing diverse objects. Therefore, a data $r$ (which is the object most different than the rest of the cluster) in $C$ is increasingly different from objects in the same cluster as $C$ changes into a larger cluster $C'$. In this case, $r$ should be split out from $C$.
% that the cluster $C$ merges with some other clusters forming a larger cluster $C'$. However, the data$r$ in $C$ is increasingly different from objects in the same cluster as $C$ changes into a larger cluster $C'$. This suggests that $C'$ should split $r$. The second scenario is that the data$r\in C$ is more similar to the objects of another cluster $C'$ compared with the objects in $C$. In this case, merging $r$ with $C'$ is a better clustering. Thus, $r$ needs to be split out from $C$ first. In both of the two scenarios, $r$ is the datawhich is different from most of the objects in the same cluster.
So the heuristic is that the objects which are most different from the other objects in the same cluster will be prioritized to split. If a cluster $C$ is predicted as $1$ by the split model, we try to split it in the following steps.
\begin{enumerate}
    \item  For each object $r\in C$, denote the inter similarity between $\{r\}$ and $C-\{r\}$ as the weight of $r$.  Rank the objects in $C$ in a decreasing order with their weights and put them in the queue {\bf Q} sequentially.  
    \item  Check the first object $r$ in {\bf Q}, evaluate if splitting $r$ out of $C$ generates a better clustering. If it does, then go to step $3$. Otherwise, remove $r$ from {\bf Q} and repeat step $2$.
    \item  Create a new cluster $C'=\{r\}$ and remove $r$ from $C$. 
\end{enumerate}

If a cluster $C$ is split, then it will generate two clusters $\{r\}$ and $C-\{r\}$. Each time, we just split one object from a cluster for two reasons. 
% First, this strategy will significantly improve the efficiency since its time complexity is $O(n)$ where $n$ is the number of the objects.  
First, the \textit {Split} model can be used to predict possible split changes many times in the latter rounds so that we need not worry that some clusters have not been split completely.
Second, we observe that most clusters split into two clusters where one of them often has a small size. This is because if new updates cause many objects splitting out from its original cluster $C$, then this suggests that $C$ does not have high cohesion, thus would have been split earlier.  
\begin{algorithm}[t]
      %\SetAlgoNoLine 
      \LinesNumbered
      \SetKwInOut{Input}{\textbf{Input}}\SetKwInOut{Output}{\textbf{Output}}
      
    \BlankLine
    $Cl_{merge} \leftarrow \emptyset$, change $\leftarrow$ false\;
    Put the clusters predicted as $1$ by merge model into $Cl_{merge}$ \;
    \While {$Cl_{merge} \neq \emptyset$}
    {dequeue $C\in Cl_{merge}$\;
    %check whether $C$ can merge with the specific cluster $C'$ in $Cl_{merge}$\;
    \If {$C$ can merge with the selected cluster $C'$ in $Cl_{merge}$}{
    merge $C$ and $C'$\; 
    remove $C$ and $C'$ from $Cl_{merge}$\;
    change $\leftarrow$ true;
    }
    \Else{ remove $C$ from $Cl_{merge}$;
    }
    }
    
    \caption{Merge Algorithm}
\label{algo:merge}
\end{algorithm}

Algorithm~\ref{algo:split} shows how the split algorithm works. The split algorithm first leverages the \textit{Split} model to predict which clusters should be split and put them in the set $Cl_{split}$ (line $2$). For each cluster $C\in Cl_{split}$, check whether $C$ be split into two clusters $r$ and $C-\{r\}$ where $r$ is selected according to the strategy presented in the split algorithm (line $5$). If yes, then it removes C from $Cl_{split}$ and makes the change status as true (line $6-8$). Otherwise, it just removes C from $Cl_{split}$ (line $11$). Thisprocess continues until $Cl_{split}$ is empty (line $3-13$).

\begin{algorithm}[h]
      %\SetAlgoNoLine 
      \LinesNumbered
      \SetKwInOut{Input}{\textbf{Input}}\SetKwInOut{Output}{\textbf{Output}}

    \BlankLine
    $Cl_{merge} \leftarrow \emptyset$, change $\leftarrow$ false\;
    Put the clusters predicted as $1$ by split model into $Cl_{split}$ \;
    \While {$Cl_{split} \neq \emptyset$}
    {dequeue $C\in Cl_{merge}$\;
    %check whether $C$ can split into two clusters $\{r\}$ and $C-\{r\}$\;
    \If {$C$ can split into two clusters $\{r\}$ and $C-\{r\}$}{
    split $C$ into $\{r\}$ and $C-\{r\}$\;
    remove $C$ from $Cl_{split}$\; 
    change $\leftarrow$ true;
    }
    \Else{ remove $C$ from $Cl_{split}$;
    }
    }
    % \For {$C\in Cl_{split}$}
    % {check whether $C$ can split into two clusters $\{r\}$ and $C-\{r\}$\;
    % }
    \caption{Split Algorithm}
\label{algo:split}
\end{algorithm}

\subsection{Full Algorithm}
%\noindent{\bf Full algorithm:}
The full algorithm of \textit{DynamicC} is shown in Algorithm \ref{algo:dynamic}.  At a high level, \textit{DynamicC} performs the merge and split algorithms alternately.  
Initially, it processes the updates according to the approach introduced in  Section~\ref{sec:initial} (line $1$). This way the algorithm starts with an initial clustering where each new or updated object is assigned to a cluster. Next, it executes the merge algorithm (line $5$) and split algorithm (line $6$).   
If some clusters are merged or split, it perform the merge and split algorithms again (line $4-7$). DyanmicC terminates when no clusters can be merged or split.

\textbf{Algorithm Properties.}
The \textit{DynamicC} algorithm has two properties. First, the \textit{DynamicC} algorithm converges because both of the merge and split algorithms are decreasing the objective score of the clustering. Since there is a minimal objective score for the clustering result and the number of different clustering results are finite, \textit{DynamicC} converges. Second, the complexity of the DynamicC is $O(2kn)g(|\mathcal{L}_{R}|)$, where $k$ is the number of iterations that DynamicC needs for converging, $n$ is the number of objects and $g(|\mathcal{L}_{R}|)$ is the time of evaluating the objective function on clustering $\mathcal{L}_{R}$. Note that for both of the merge algorithm and split algorithm, each of them takes $O(n)g(|\mathcal{L}_{R}|)$ time for each round. Therefore, \textit{DynamicC} algorithm takes $O(2n)g(|\mathcal{L}_{R}|)$ time. In practice, the number of clusters that are predicted as ``merge'' or ``split'' is much smaller than $n$.

\begin{algorithm}[h]
      %\SetAlgoNoLine 
      \LinesNumbered
      \SetKwInOut{Input}{\textbf{Input}}\SetKwInOut{Output}{\textbf{Output}}
      
      \Input{
          Similarity matrix for all the objects; \\
          ~Initial clustering and the updates; \\
          ~Merge and split model: $\mathcal{M_M}$, $\mathcal{S_M}$.}
    \Output{
        The new clustering  $Cl_{new}$.}
    \BlankLine
    Process the updates (see the approach introduced in the beginning of Section \ref{sec:initial}) \;
    $Cl_{merge} \leftarrow \emptyset$, $Cl_{split} \leftarrow \emptyset$\;
    
    change $\leftarrow$ true\;
    \While {change}{
    Execute merge algorithm (Algorithm~\ref{algo:merge})\;
    Execute split algorithm (Algorithm~\ref{algo:split})\;
    }
  \Return {$Cl_{new}$} \;
    \caption{DynamicC Full Algorithm}
\label{algo:dynamic}
\end{algorithm}

\section{Experiment}
We present the experimental results on four real-world datasets and a synthetic dataset. The results show that our method successfully speeds up clustering compared to the underlying batch algorithm while maintaining a clustering accuracy close to the ones obtained by the batching algorithm.
We compare with three methods: a batching solution, a naive incremental solution, and a state-of-the-art incremental solution.

\begin{table*}
\setlength{\abovecaptionskip}{0.2cm}
\setlength{\belowcaptionskip}{-0.cm}
\caption {Experimental settings on the datasets}
\label{dataset:setting}
\centering
\begin{tabular}{|c|c|c|c|c|c|c|}
%{|m{1.6cm}|m{1.4cm}|m{1.7cm}|m{1.8cm}|m{1.8cm}|m{1.8cm}|}
\hline
\bfseries Dataset &  \bfseries similarity/distance    & \bfseries \# of initial objects    & \bfseries \# of final objects & \bfseries data type \\\hline
{\bfseries Cora}  & Jaccard~\cite{niwattanakul2013using} & 279  & 1879   & textual and numerical \\\hline
{\bfseries Music Brain (Music)}  & Cosine Trigram similarity~\cite{nentwig2018incremental} & 4K & 15375 & textual  \\\hline
{\bfseries Amazon Access Samples (Access)}  & Euclidean distance & 1K   &20208    & numerical   \\\hline
{\bfseries 3D Road Network (Road)}  & Euclidean distance &100K    &344768   & numerical  \\\hline
{\bfseries Synthetic}  & Levenshtein~\cite{yujian2007normalized} and Jaccard & 10K & 43K  & textual and numerical \\\hline
\end{tabular}
\end{table*}

\subsection{Setup}
\noindent{\bf Datasets:} We experimented on five datasets. The first real-world dataset, Cora \cite{Cora}, contains $1879$ publication objects and has been widely used for  clustering. 
The second real-world dataset, Music Brainz (Music for short)~\cite{Music}, is a collection about music datasets. It contains $19375$ objects.
The third real-world dataset, Amazon Access Samples (Access for short) \cite{Amazon}, is an anonymized sample of access provisioned within the company. It contains 30000 instances.
The forth real-world dataset, 3D Road Network (Road for short) \cite{RoadNetwork}, includes 3D road network with highly accurate elevation information. It contains 434874 instances.
The fifth dataset is a synthetic dataset. We use the $\textit {Febrl}$ \cite{Febrl} data generator to generate the synthetic dataset. This tool generates a dataset with a number of original and duplicate objects specified by the user. The distribution of the duplicate objects is also set by the user. To experiment with different distributions, we generate three datasets with uniform, poisson and zipf distributions for duplicates. In our experiments with Febrl, we start with a dataset consisting of 18K original objects and $30K$ duplicate objects.

Table\ref{dataset:setting} summarizes information about the three datasets we use in this section. 
Our workloads mimic dynamic processes including three kinds of updates: Adding new objects (Add for short), Removing existing objects (Remove for short) and Updating existing objects (Update for short). We explain the details in later sections for each dataset.

% \begin{table}[h]
% \centering
% \begin{tabular}{|c|c|c|c|}
% %{|m{1.6cm}|m{1.4cm}|m{1.7cm}|m{1.8cm}|m{1.8cm}|m{1.8cm}|}
% \hline
% \bfseries Statistics/Datasets &  \bfseries  Cora   & \bfseries Music Brainz & \bfseries Synthetic\\\hline
% {\bfseries \# of objects}  & 1879 & 19375 & 48K \\\hline
% {\bfseries Sim $\in$[0.9, 1.0]}  & 29983 & 10230 & 164163 \\\hline
% {\bfseries Sim $\in$[0.85, 0.9)}  & 58 & 5384 & 69135\\\hline
% {\bfseries Sim $\in$[0.8, 0.85)}  & 738 & 4384 & 109741\\\hline
% {\bfseries Sim $\in$[0.75, 0.8)}  & 2567 & 3604 & 49084\\\hline
% {\bfseries Sim $\in$[0.7, 0.75)}  & 1169 & 2630 & 18441\\\hline
% {\bfseries Avg #neighbors}  & 18.8 & 1.4 & 8.6\\\hline
% \end{tabular}
% \caption{Characteristics of the datasets used in the evaluation}
% \label{table:statistics}
% \end{table}
% \vspace{-20pt}

\noindent{\bf Implementations:} The experimental environment is a computer with Quad-Core Intel Core i5  processor,  8GB  memory, running macOS Catalina. The ML model is implemented in Python with scikit-learn~\cite{pedregosa2011scikit} library.  The core algorithm is implemented in Java and we execute the python scripts in the java programs. We use the Logistic Regression Model as the default ML model for comparison unless we mention otherwise.

\noindent{\bf Comparison:}
We evaluate our method on three kinds of clustering methods of increasing complexity: DBSCAN, $k$-means and DB-index. We briefly describe the three kinds of clustering problems and then show the reasons about why we evaluate based on the three kinds of clustering. 
\begin{itemize}
\item {\bf Density-based clustering:} The idea of density-based methods is to continue growing as long as the number of objects in the neighborhood exceeds some thresholds.

\item {\bf $k$-means:} k-means clustering \cite{macqueen1967some} is a classical unsupervised learning clustering method which aims to partition objects into $k$ clusters in which each object belongs to the cluster with the nearest mean. Finding the optimal solution for k-means clustering is computationally difficult (NP-hard). Although there are many heuristic algorithms~\cite{lloyd1982least, pelleg1999accelerating} for $k$-means clustering, we use a more robust batch algorithm to test the effectiveness of our dynamic method. 
% To the best of our knowledge, there is no existing dynamic algorithm for k-means clustering.

\item {\bf DB-index:} Davies-Bouldin index (DB-index) \cite{gruenheid2014incremental} was originally defined for Euclidean space~\cite{davies1979cluster}. With some adjustment for the definition of distance, it is adapted to the problem of record linkage~\cite{guo2010record,gruenheid2014incremental}. Unlike $k$-means clustering, DB-index does not assume the number of clusters a priori. Finding the optimal solution for DB-index is also intractable.
Compared to correlation clustering \cite{bansal2004correlation}, DB-index clustering is more challenging because it has no special properties for optimizing clustering~\cite{gruenheid2014incremental}. 
\end{itemize}

% \Change{Maybe make a clarification about the work of Gan et.al.~\cite{gan2017dynamic} to explain why we do not evaluate their work in the experiments.}

We choose the three clustering problems for two reasons. First, they are representative of difference clustering approaches: density-based, partitioning and objective-based clustering. Second, the complexity of the three clustering methods increases and there is no straightforward dynamic strategy for them.
For the above three kinds of clustering problems, we implement two batch algorithms: {\bf DBSCAN} and {\bf Hill-climbing}. Also, we implement one naive incremental method {\bf Naive}, and one state-of-the-art incremental method {\bf Greedy}~\cite{gruenheid2014incremental} which is the most similar to our work. 

\begin{figure*}[htb]
\centering
\subfigure[\Change{Operations for each dataset}]{
    \begin{minipage}[h]{0.62\textwidth}
    \centering
    \includegraphics[width=1\textwidth, height=4cm]{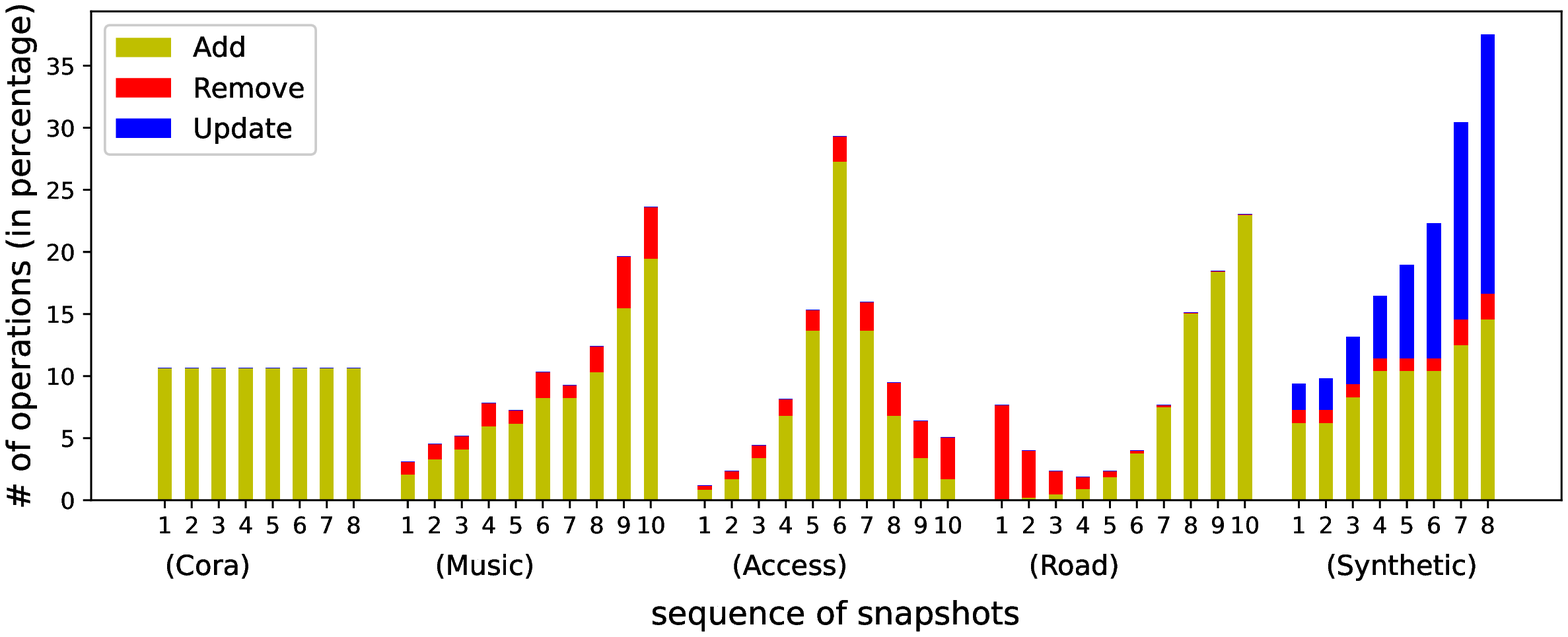}
    %\label{subfig:change}
    \end{minipage}
    }
% \subfigure[Add operations for each dataset]{
%     \begin{minipage}[h]{0.31\textwidth}
%     \centering
%     \includegraphics[width=1\textwidth]{pics/expe/change_add.eps}
%     \end{minipage}
%     }
\subfigure[Re-clustering latency on Access]{
    \begin{minipage}[h]{0.31\textwidth}
    \centering
    \includegraphics[width=1\textwidth]{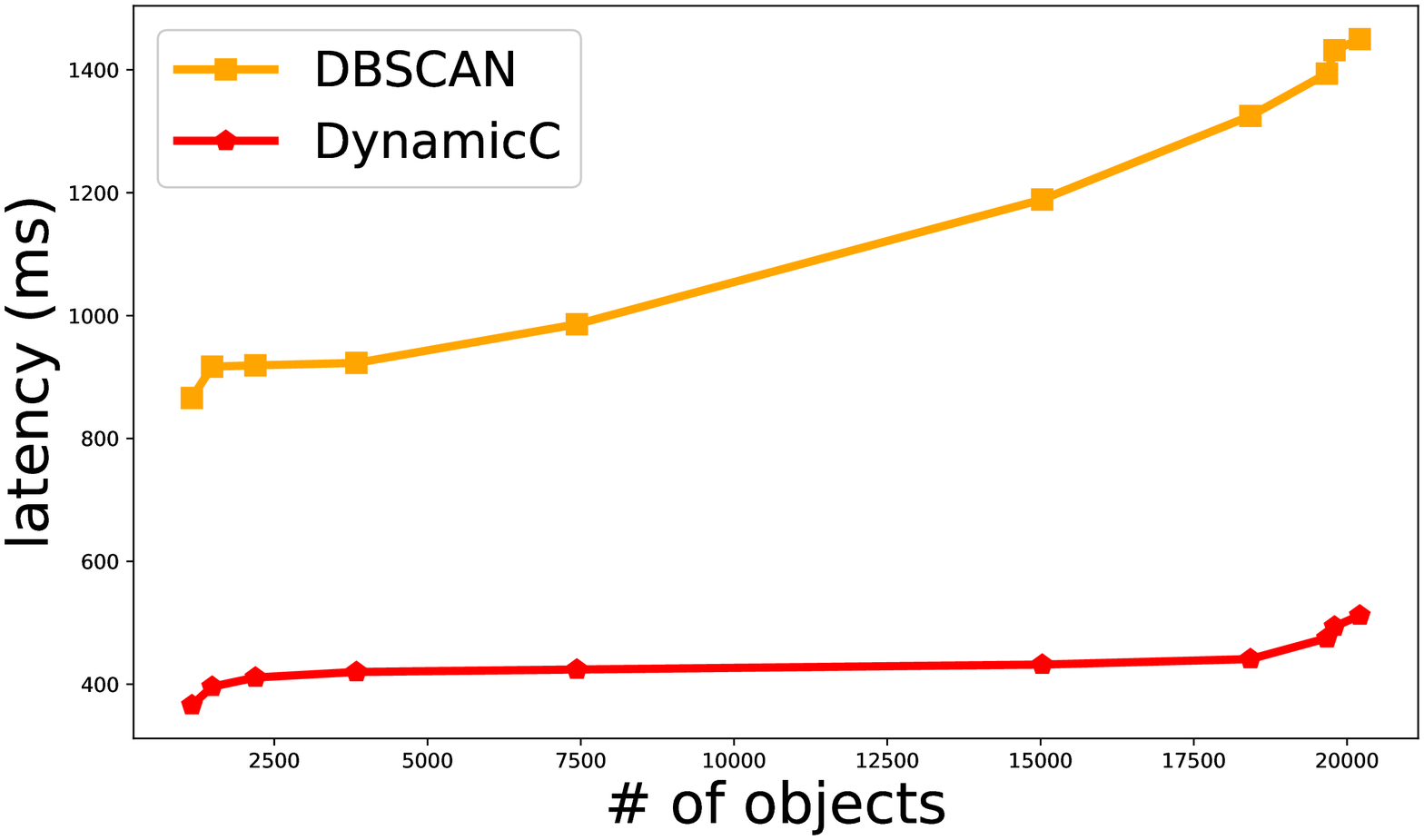}
    %\label{subfig:access_time}
    \end{minipage}
    }
\subfigure[Re-clustering latency on Road]{
    \begin{minipage}[h]{0.31\textwidth}
    \centering
    \includegraphics[width=1\textwidth]{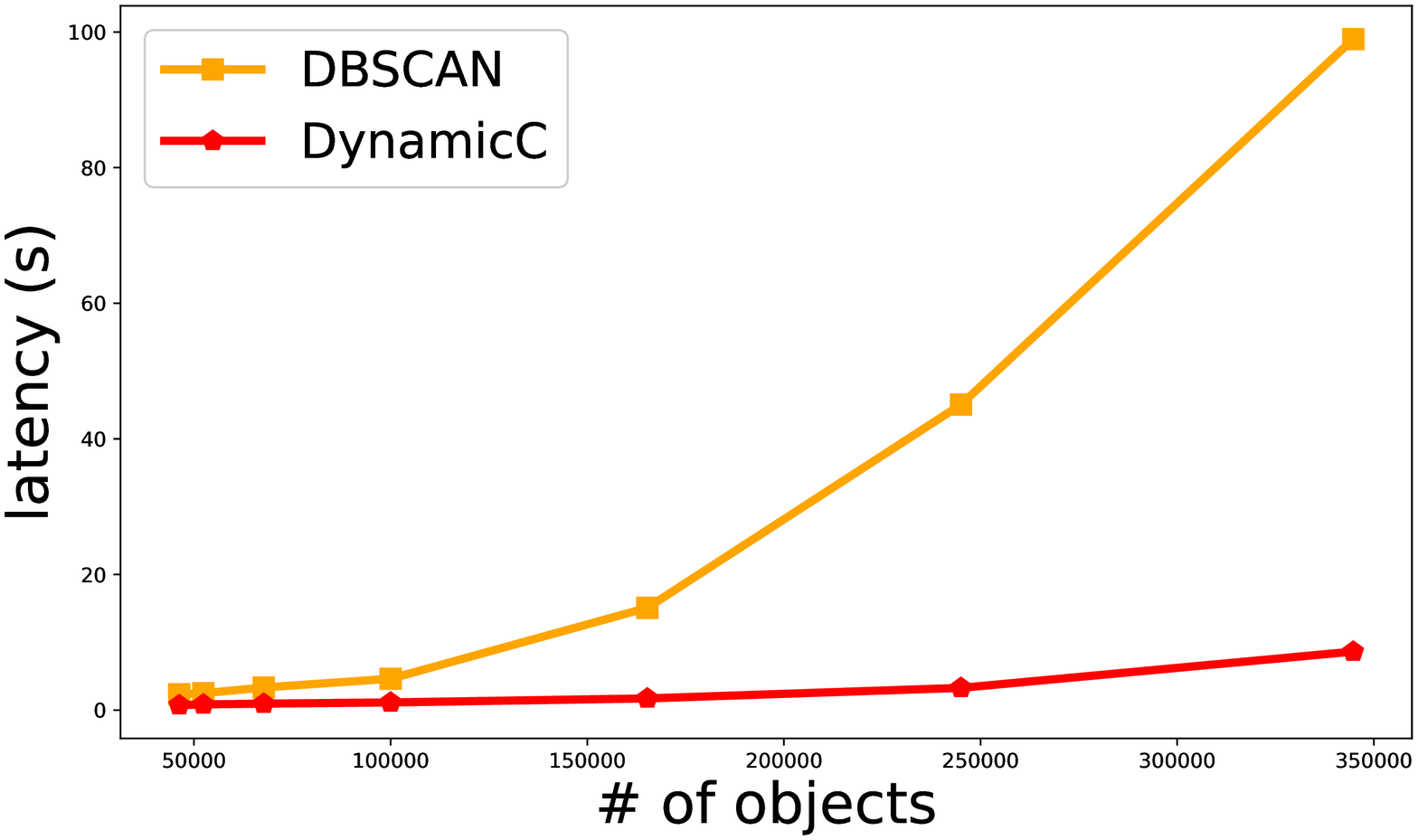}
    %\label{subfig:road_time}
    \end{minipage}
    }
\subfigure[Square root of objective score on Road (Lower values better)]{
    \begin{minipage}[h]{0.31\textwidth}
    \centering
    \includegraphics[width=1\textwidth]{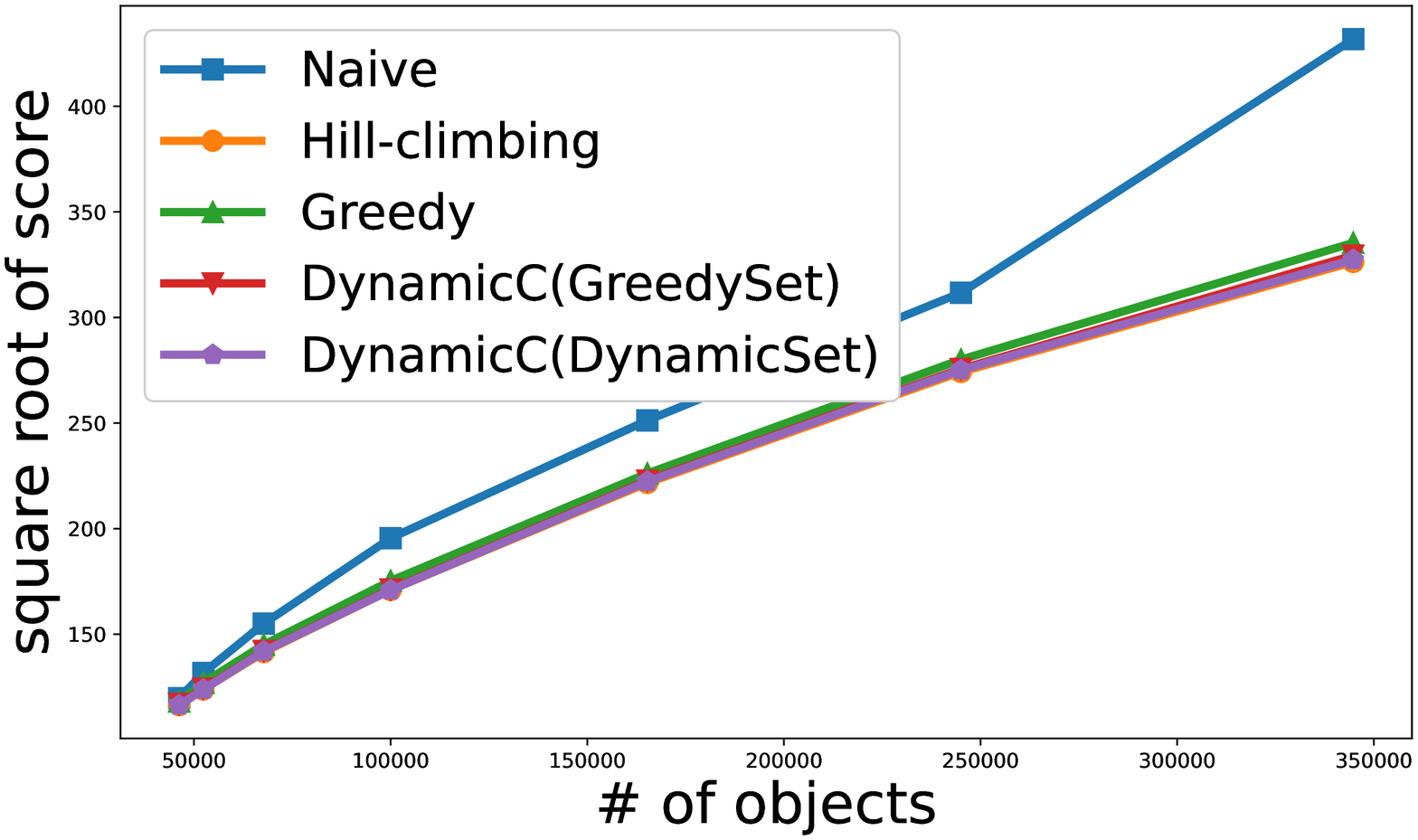}
    %\label{subfig:square}
    \end{minipage}
    }
% \subfigure[Remove and Update operations for each dataset]{
%     \begin{minipage}[h]{0.31\textwidth}
%     \centering
%     \includegraphics[width=1\textwidth]{pics/expe/change_remove.eps}
%     \end{minipage}
%     }
\subfigure[Re-clustering latency on Road (the curve of Hill-climbing is omitted since it takes more than $3$ hours on average)]{
    \begin{minipage}[h]{0.31\textwidth}
    \centering
    \includegraphics[width=1\textwidth]{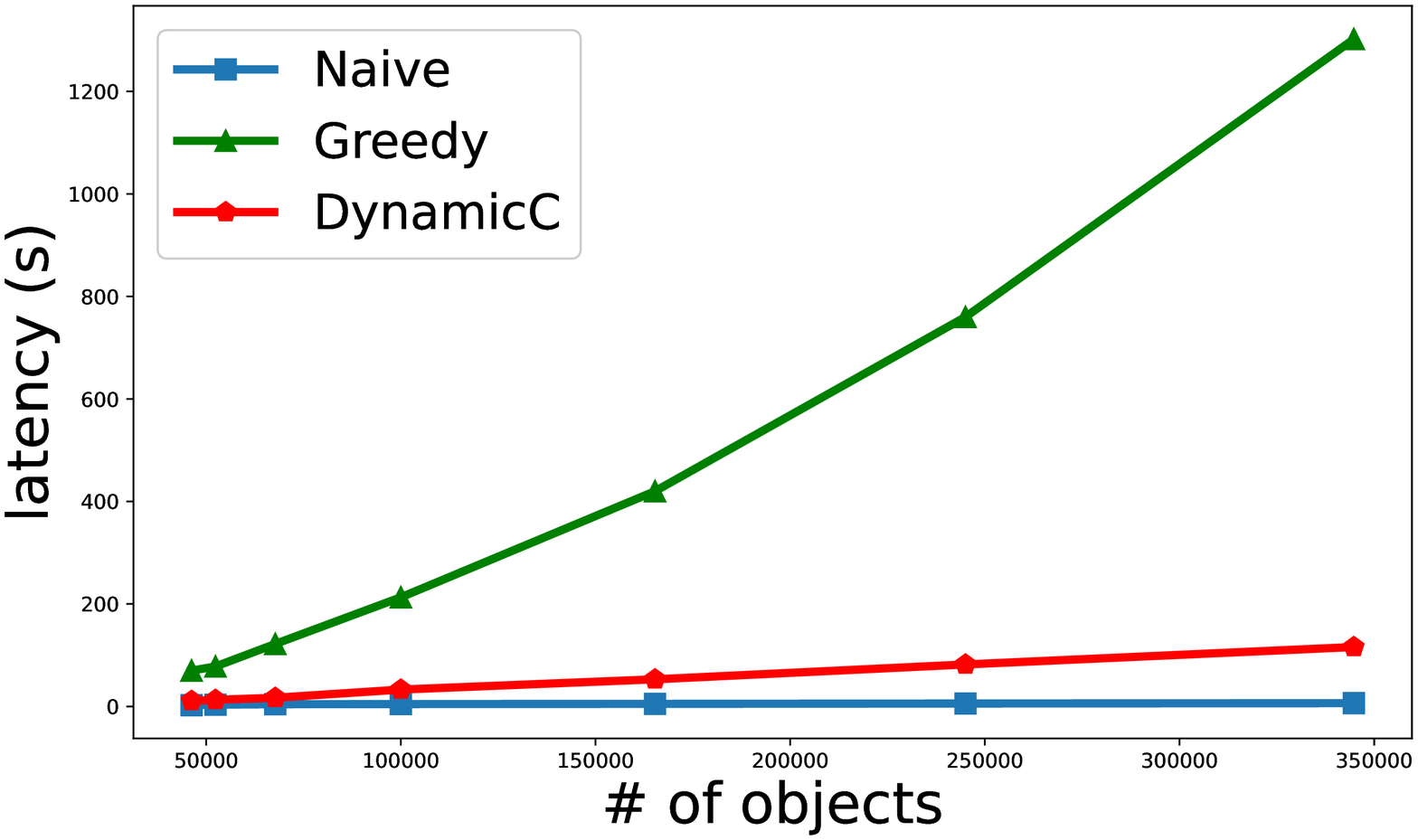}
    \end{minipage}
    %\label{subfig:road_latency}
    }
\caption{Dynamic process on the datasets and the results for density-based and k-means clustering}
\label{fig:DB_K}
\end{figure*}

\begin{itemize}
\item {\bf DBSCAN:} DBSCAN \cite{ester1996density} uses a minimum density level estimation, based on a
threshold for the number of neighbors, minPts, within the radius $\varepsilon$. 
Objects with more than minPts neighbors within this radius 
are taken as a core point. 

\item {\bf Hill-climbing:} This is a general batch algorithm which can be used for any objective function based clustering method \cite{russell2002artificial}. It examines all immediate neighbors (potential migrations) and selects the clustering update providing the highest improvement. This method is applicable to many other clustering methods as it does not require any ad-hoc rules or properties of the clustering methods.

\item {\bf Naive:} This is the baseline incremental algorithm. It compares each new object with existing clusters and then assign an object to the closest cluster or a new cluster. 
% (we choose the one with the maximal inter-similarity in the experiments).
This method does not compute the objective score for the clustering. Its decisions are only based on heuristics such as similarity threshold.  

\item {\bf Greedy:} This method is proposed in \cite{gruenheid2014incremental}. It uses three operators to determine a candidate clustering which makes it able to terminate in polynomial time.
\end{itemize}

{\bf DBSCAN} is only applied for density-based clustering. {\bf Hill-climbing}, {\bf Naive} and {\bf Greedy} are applied to $k$-means and DB-index clustering in the following comparisons.

The accuracy of {\bf DynamicC} depends on the clustering that it starts with. For this reason, we provide evaluations with two scenarios: (1)~\textit{GreedySet} scenarios: in this case, the starting point for each round of updates is set to the results of {\bf Greedy} from the previous clustering so that {\bf Greedy} and {\bf DynamicC} start with the same clustering; (2)~\textit {DynamicSet} scenarios: in this case, the starting point is the dynamic clustering results from the previous clustering.
In a real deployment, DynamicSet matches the typical execution, while GreedySet matches the case when the underlying batch algorithm is run occasionally.
To generate the initial cluster evolution for DynamicC, we use the Hill-climbing method.

For correlation clustering, its property has been well studied  in \cite{gruenheid2014incremental}. The {\bf Greedy} approach proposed in \cite{gruenheid2014incremental} shows its effectiveness with the correlation clustering and most of the experimental results are based on the correlation clustering. However, as described in \cite{gruenheid2014incremental}, DB-index clustering is more challenging because it does not have some desirable properties that can be used for optimization. 
 
\begin{figure*}[htb]
\setlength{\abovecaptionskip}{0.cm}
\setlength{\belowcaptionskip}{-0.2cm}
\centerline{\psfig{figure=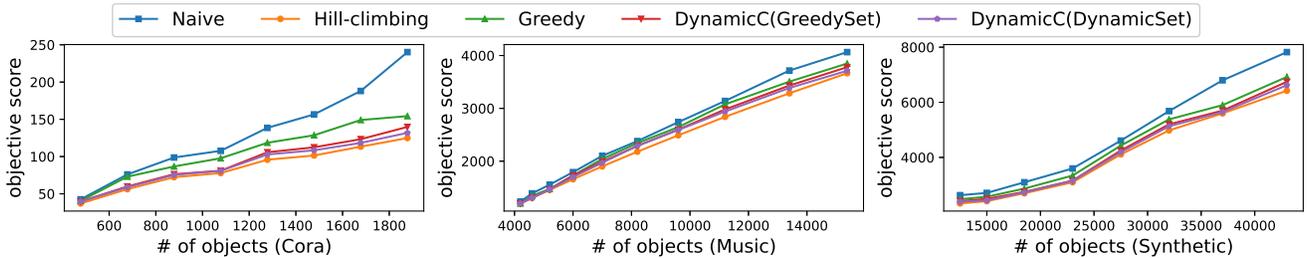, width=0.99\textwidth} }
\caption{Quality on Cora, Music and Synthetic datasets (Lower values are better)}
\label{fig:db_score}
\end{figure*}
\begin{figure*}[htb]
\setlength{\abovecaptionskip}{0.cm}
\setlength{\belowcaptionskip}{-0.2cm}
\centerline{\psfig{figure=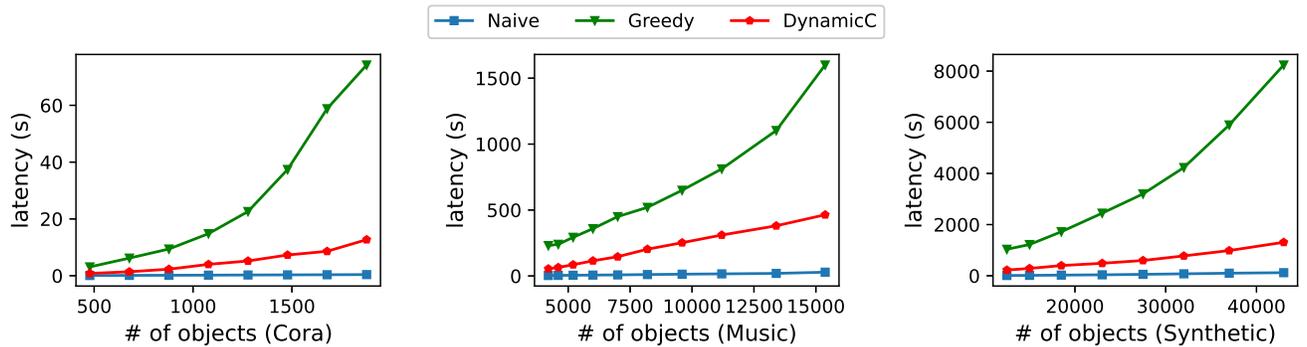, width=0.99\textwidth} }
\caption{Re-clustering latency on Cora, Music and Synthetic datasets (the curve of Hill-climbing is omitted since it takes more than $4$ hours on average for each dataset)}
\label{fig:db_time}
\end{figure*}

\noindent{\bf Measurement:} We evaluate our method in terms of latency and quality of clustering results. For latency, we report the average execution time of 20 runs of the experiments. In each round, we focus on the latency of re-clustering as well as the model re-training time. In our experiments, the reported latency is of both the re-training and re-clustering. 
% It turns out that the latency of re-training the model is negligible compared to the time to re-cluster. 
% Computing the similarity among objects only takes a little time compared to clustering in our case. There are many blocking techniques that can be used to improve the efficiency of pairwise similarity reference, however such work is orthogonal to our proposal.  
For measuring the quality of clustering, we report the objective score and pair counting $F1$ measure \cite{achtert2012evaluation}, precision, recall, purity and inverse purity. As the goal of dynamic clustering is to get the similar clustering result with that of the batch methods, we take the clustering result of the batch method as the ground truth.

\subsection{Quality and Performance}
More details about the experimental settings on each dataset is presented in Table~\ref{dataset:setting}. To mimic the dynamic process, we randomly select some objects from the original dataset as the initial dataset in the beginning. Next, at each snapshot (round), we randomly add, remove or update some objects and start re-clustering. The number of added, removed and updated objects for each dataset is shown in Figure~\ref{fig:DB_K}(a).
% The numbers are represented with its ratio to the size of the original dataset. 
%
% To mimic a dynamic process, we randomly select some objects from the original dataset as the initial dataset. After that, we take $8$ or $10$ rounds of updates.
We only perform the Update operations on the Synthetic dataset by the Febrl Generator since Febrl allows us to generate similar objects as well as do modifications to attribute values. Update operation can be factorized into Remove and Add operation, so it is fine with only Remove and Add operations.
Since DBSCAN and k-means often work for numerical data, we only report the results on Access and Road datasets for them. 
And we report the results on Cora, Music and Synthetic datasets for DB-index clustering.

\vspace{-10pt}
\subsubsection{\bf Results for DBSCAN}
Although {\bf DynamicC} is designed with objective function-based clustering in mind, we test {\bf DynamicC} to show that our method can be augmented with many other clustering methods with minor changes. 
Since DBSCAN has no objective function, we judge if the clustering is good or not by checking whether the relevant previous core points are stable. 

In DBSCAN, changes in the clusters are based on two parameters: minPts and $\varepsilon$ for each new or old object. We notice that {\bf DynamicC} does not work well when the objects in the final cluster are very sparse (i.e. objects in a cluster are not compact). Different minPts and $\varepsilon$ settings can significantly affect the time cost and clustering result for DBSCAN. We report the average F1 and latency with 10 groups of parameters.  The average F1 scores on Access and Road are $0.988$ and $0.976$, respectively, across all the snapshots for {\bf DynamicC}. {\bf DynamicC} does not achieve the same results as DBSCAN since we do not check each object and do not follow its exact rules. However, {\bf DynamicC} can still get good F1 score faster as shown in Figure~\ref{fig:DB_K}(b) and (c).
In addition, DynamicC does not always get better clustering result than manually designed methods like \cite{DBLP:conf/vldb/EsterKSWX98} (while \cite{gan2017dynamic} focuses on $\rho$ approximate DBSCAN and use another metric to evaluate their method instead of F1). However, DynamicC saves around $40\%$ to $60\%$ time while reaching F1 scores that are close to the optimal.

\vspace{-5pt}
\subsubsection{\bf Results for $k$-means}
The F1 scores of most snapshots are $1$ or very close to $1$ on Access and Road datasets. This can also be seen from 
Figure~\ref{fig:DB_K}(d) which shows the square root of objective score on Road dataset (the results on Access dataset are similar to that of Road dataset). {\bf Naive} method gets worse objective scores when more updates arrive. But the other methods get almost the same objective score with that of the batch algorithm. The reason is that when the number of clusters is fixed (i.e. the value of $k$), it is easy for the other methods to converge to a stable clustering which is often a near optimal clustering. However, {\bf DynamicC} significantly reduces the latency as shown in Figure~\ref{fig:DB_K}(e).

\vspace{-5pt}
\subsubsection{\bf Results for DB-index}
The quality evaluation of five approaches is shown in Figure~\ref{fig:db_score} for DB-index clustering. The experimental results on three datasets are similar.
The quality of {\bf Naive} method decreases when the number of objects increases to be the worst out of the other methods. 
The reason for this is that the ``merge-only" strategy applied in {\bf Naive} can not work well when the clustering structure changes.

{\bf Hill-climbing} achieves the best objective score among the five approaches as it searches for a better clustering in a brute-force way. {\bf Greedy} achieves better score than {\bf Naive}, but gets worse score than {\bf DynamicC}. 
{\bf DynamicC} reaches better objective score in DynamicSet than the GreedySet, showing that that the quality of the initial clustering can make {\bf DynamicC} obtain better objective score.  Table~\ref{table:DB-index_F1} shows the F1 measures of {\bf Naive}, {\bf Greedy} and {\bf DynamicC} (we only show the results of $5$ out of $8$ (for Cora and Synthetic) or $10$ snapshots (for Music) due to space limit). {\bf Naive} gets worse and worse F1 measure when the number of updates increases. {\bf DynamicC} gets a little higher F1 measure than that of {\bf Greedy} can obtain similar result with that of the batch algorithm since its F1 measure is very close to $1$.
Table~\ref{table:DB-index_other} shows the performance of three methods with precision, recall, purity~\cite{zhao2001criterion} and inverse purity~\cite{amigo2009comparison}. Due to space limit, we just show the results based on the last round for each dataset. As can be seen from the table, DynamicC can get the best precision and recall. And because purity and inverse purity focus on the cluster
with maximum precision and recall respectively, DynamicC can also get the best performance in terms of purity and inverse purity. We do observe that DynamicC occasionally gets lower precision but higher recall than Greedy during the whole dynamic process. Nevertheless, DynamicC gets best F1 scores in most cases which can be seen from Table \ref{table:DB-index_F1}.
%(i.e. with multiple rounds). 

For latency, the results are shown in Figure~\ref{fig:db_time} (the points of each line corresponds to the snapshots during the dynamic process).  {\bf DynamicC} has similar execution time in both GreedySet and DynamicSet scenarios. So, we take their average execution time as the time cost of {\bf DynamicC}. {\bf DynamicC} significantly reduces the time cost especially when the scale of the dataset is large. Also, Figure~\ref{fig:db_time} demonstrates that the latency of {\bf Greedy} increases significantly compared to Naive and DynamicC when the dataset becomes larger.
We also observe that {\bf DynamicC} saves significantly more overhead than {\bf Greedy} on the Synthetic dataset when the dataset becomes larger. This indicates that {\bf DynamicC} can better address complex data structure as the average number of similar neighbor objects in the Synthetic dataset is larger than that of Cora and Music Brainz datasets. 

In summary, {\bf DynamicC} dramatically decreases latency compared to {\bf Greedy} and {\bf Hill-climbing} while retaining or approaching the quality of clustering even when the dataset is large.

\begin{table}
\setlength{\abovecaptionskip}{0.1cm}
\setlength{\belowcaptionskip}{-0.cm}
\caption{F1 measure for DB-index clustering}  
\label{table:DB-index_F1}
\centering
\scriptsize
\renewcommand\arraystretch{1.2}
\begin{tabular}{|p{1.8cm}|c|p{0.6cm}|p{0.6cm}|p{0.6cm}|p{0.6cm}|p{0.6cm}|}
\hline
\bfseries Dataset/Snapshot & \bfseries  & \bfseries 1  & \bfseries 2 & \bfseries 3 & \bfseries 4 & \bfseries 5  \\\hline

\multirow{2}{*}{\bfseries Cora}
& Naive & 0.943 & 0.912 & 0.908  & 0.871  & 0.843    \\\cline{2-7}
& Greedy  & 0.998 & 0.985 & 0.984  & 0.981  & 0.981 \\\cline{2-7}
& DynamicC  & \textbf{1} & \textbf{0.988} &  \textbf{0.991}  & \textbf{0.983}  & \textbf{0.984}       \\\hline

\multirow{2}{*}{\bfseries Music}
& Naive & 0.982 & 0.976 & 0.963  & 0.945  & 0.932    \\\cline{2-7}
& Greedy  & 1 & 0.991 & 0.987  & 0.986  & 0.989 \\\cline{2-7}
& DynamicC  & \textbf{1} & \textbf{0.996} &  \textbf{0.994}  & \textbf{0.991}  & \textbf{0.993}       \\\hline

\multirow{2}{*}{\bfseries Synthetic}
& Naive & 0.931 & 0.871 & 0.864  & 0.831  & 0.815    \\\cline{2-7}
& Greedy  & 0.995 & 0.985 & \textbf{0.991}  & 0.984  & 0.979 \\\cline{2-7}
& DynamicC  & \textbf{0.998} & \textbf{0.997} &  0.989  & \textbf{0.994}  & \textbf{0.992}       \\\hline
\end{tabular}
\end{table}

% \vspace{-1pt}
\begin{table}
\setlength{\abovecaptionskip}{0.1cm}
\setlength{\belowcaptionskip}{-0.cm}
\caption{Other metrics for DB-index clustering}
\label{table:DB-index_other}
\centering
\scriptsize
\renewcommand\arraystretch{1.2}
\begin{tabular}{|p{1.8cm}|c|p{0.8cm}|p{0.8cm}|p{0.8cm}|p{0.8cm}|}
\hline
\bfseries Dataset/Metrics & \bfseries  & \bfseries precision  & \bfseries recall & \bfseries purity & \bfseries inverse purity   \\\hline

\multirow{2}{*}{\bfseries Cora}
& Naive  & 0.884 & 0.806  & 0.914  & 0.842    \\\cline{2-6}
& Greedy   & 0.992 & 0.970  & 0.994  & 0.984 \\\cline{2-6}
& DynamicC  & \textbf{0.996} &  \textbf{0.972}  & \textbf{0.997}  & \textbf{0.988}       \\\hline

\multirow{2}{*}{\bfseries Music}
& Naive  & 0.913 & 0.952  & 0.943  & 0.976    \\\cline{2-6}
& Greedy & 1 & 0.978  & 1  & 0.992 \\\cline{2-6}
& DynamicC  & \textbf{1} &  \textbf{0.986}  & \textbf{1}  & \textbf{0.994}       \\\hline

\multirow{2}{*}{\bfseries Synthetic}
& Naive  & 0.835 & 0.796  & 0.879  & 0.861    \\\cline{2-6}
& Greedy   & 0.987 & 0.971  & 0.976  & 0.986 \\\cline{2-6}
& DynamicC  & \textbf{0.990} &  \textbf{0.994}  & \textbf{0.999}  & \textbf{0.992}       \\\hline
\end{tabular}
\end{table}

\subsection{ML Model Evaluation}
\label{sec:robustness}
In this section, we evaluate the effectiveness of ML models in DynamicC and their retraining time. 

Table~\ref{table:model} shows the accuracy and recall of Logistic Regression, SVM and Decision Tree models. When the number of training samples increases, the accuracy and recall of the three models all reach satisfactory values. The training time of these models is negligible as it is less than $1$ second for all the three models when the number of samples is less than 20K. The ML models do affect the latency of DynamicC. The larger precision and recall of the models, the less latency of DynamicC.  Nevertheless, DynamicC gets almost the same F1 score with different ML models finally.  This is because DynamicC chooses a proper $\theta$ value (described in Section~\ref{sec:guarantee}) to make sure that the model will get high recall for new objects.

We evaluate the performance of Logistic Regression model in terms of the fraction of training data with respect to the three datasets. Table~\ref{table:model-lg} shows the results with default parameter settings. For each round of evaluation, 20\% of the whole objects are added for model evaluation. 
Initially, the model has low accuracy and recall since it has little information about the clustering result. But the model becomes experienced with whether the clustering is good or not when more samples are provided. The accuracy of the models will reach stability when sufficient samples are fed.
This indicates that these models are effective when more samples are provided.

\begin{table}
\caption{Evaluation for different ML models on Cora dataset (the training time is less than $1$s for all the three models when the number of samples is 20K)} 
\label{table:model}
\setlength{\abovecaptionskip}{0.1cm}
\setlength{\belowcaptionskip}{-0.cm}
\scriptsize
\renewcommand\arraystretch{1.2}
% \begin{tabular}{|p{1.8cm}|c|p{0.6cm}|p{0.6cm}|p{0.6cm}|p{0.6cm}|p{0.6cm}|}
\begin{tabular}{|c|c|c|c|c|c|c|}
\hline
\makecell[c]{\bfseries \# of new objects/ \\ \bfseries (\# of training samples)} & \bfseries  & \bfseries \makecell[c]{200\\(97)}  & \bfseries \makecell[c]{ 400\\(248)} & \bfseries \makecell[c]{600\\(420)} & \bfseries \makecell[c]{800\\(745)} & \bfseries \makecell[c]{1000\\(1077)}\\\hline

\multirow{2}{*} {\bfseries Logistic Regression} 
& Accuracy  & 0.77 & 0.82 & 0.88  & 0.90  & 0.93 \\\cline{2-7}
& Recall & 0.25 & 0.98 & 1.0  & 1.0  & 1.0      \\\hline

\multirow{2}{*}{\bfseries SVM}
& Accuracy & 0.77 & 0.81 & 0.87  & 0.89  & 0.92 \\\cline{2-7}
& Recall & 0.25 & 0.95 & 0.96  & 1.0  & 1.0   \\\hline
\multirow{2}{*}{\bfseries Decision Tree}
&  Accuracy & 0.86 & 0.76 & 0.86  & 0.93  & 0.95 \\\cline{2-7}
&  Recall & 0.75 & 0.80 & 0.97  & 0.96  & 1  \\\hline                        
\end{tabular}
\end{table}

\begin{table}
\caption{The performance of Logistic Regression with default settings on three datasets}
\centering
\setlength{\abovecaptionskip}{0.1cm}
\setlength{\belowcaptionskip}{-0.cm}
\label{table:model-lg}
\scriptsize
\renewcommand\arraystretch{1.2}
% \begin{tabular}{|p{1.8cm}|c|p{0.6cm}|p{0.6cm}|p{0.6cm}|p{0.6cm}|p{0.6cm}|}
\begin{tabular}{|c|c|c|c|c|c|c|}
\hline
\makecell[c]{\bfseries fraction of training samples}  & \bfseries  & \bfseries \makecell[c]{5\%}  & \bfseries \makecell[c]{10\%} & \bfseries \makecell[c]{20\%} & \bfseries \makecell[c]{40\%} & \bfseries \makecell[c]{80\%}\\\hline
\multirow{2}{*}{\bfseries Cora} & Accuracy  & 0.62 & 0.74 & 0.83  & 0.90  & 0.98  \\\cline{2-7}
                                    & Recall & 0.15 & 0.18 & 0.98  & 1.0  & 1.0       \\\hline

\multirow{2}{*}{\bfseries Music}& Accuracy & 0.84 & 0.87 & 0.94  & 0.96  & 0.97 \\\cline{2-7}
& Recall & 0.56 & 0.93 & 1.0  & 1.0  & 1.0   \\\hline

\multirow{2}{*}{\bfseries Synthetic}&  Accuracy & 0.73 & 0.85 & 0.88  & 0.89  & 0.93 \\\cline{2-7}
                                &  Recall & 0.47 & 0.81 & 0.92  & 0.95  & 0.98  \\\hline                        
\end{tabular}
\end{table}

\section{Conclusion}
In this paper, we tackle the problem of efficient clustering. We observe that existing solutions incur a significant overhead to achieve good quality. To overcome this overhead, we propose DynamicC, a machine learning-based solution that can be augmented with existing batching solutions for clustering. 
% DynamicC trains a model that learns the patterns of how clusters evolve using the underlying batching algorithm. Then, the model is used to make clustering decisions that are faster than batching solutions and similar in accuracy.
DynamicC advances the state of incremental clustering by proposing a more general dynamic system model where objects are added, deleted, and modified. Furthermore, DynamicC does not make assumptions about the underlying batching mechanisms which allows it to be augmented with a wider range of batching solutions unlike existing incremental methods. Our evaluation shows that DynamicC can achieve a significant performance improvement compared to other methods while retaining their quality.

Our future directions based on this work include applying the proposed efficient clustering methods to real problems. In particular, DynamicC is suitable for environments where there is a need for fast response from a machine learning model.
% fast ml in path
This includes problems that utilize machine learning in the path of execution of storage and transactional solutions~\cite{gazzaz2021croesus,gazzaz2022croesus,kargar2021extending,kargar2021predict,gazzaz2019collaborative, gu2019improving}.
% edge-cloud
This also includes problems where there is an asymmetry between resources, such as edge and cloud environments~\cite{nawab2021wedgechain,mittal2021coolsm}.
%blockchain
Finally, DynamicC allows deriving smaller models from bigger ones which can be useful in blockchain decentralized applications where there is a need to reduce the storage and processing footprint of applications due to cost and performance overhead~\cite{aslani2020rethinking,abadi2020anylog,alaslani2019blockchain,nawab2019blockplane}

\section{Acknowledgments}

This research is supported in part by the NSF under grant CNS-1815212 and a gift from Facebook.

\bibliographystyle{ACM-Reference-Format}
\bibliography{paperBib}

%% Please note that all the content must fit within the page limits, including any appendices.
%\appendix
% ...

\end{document}